\documentclass[a4paper,11pt]{article}
\pdfoutput=1 
\usepackage{jinstpub} 
\usepackage{natbib} 

\title{\boldmath The Multi-Blade Boron-10-based neutron detector performance using a focusing reflectometer}

\author[a,b]{G. Mauri}
\author[a]{, I. Apostolidis}
\author[c]{, M. J. Christensen}
\author[d]{, A. Glavic}
\author[a,e]{, C. C. Lai}
\author[a]{, A. Laloni}
\author[f,a]{, F. Messi}
\author[a]{, A. Lindh Olsson}
\author[a]{, L. Robinson}
\author[d]{, J. Stahn}
\author[a]{, P. O. Svensson}
\author[a,g]{, R. Hall-Wilton}
\author[a,1]{, F. Piscitelli \note{Corresponding author.}}

\affiliation[a]{European Spallation Source ERIC (ESS), P.O. Box 176, SE-22100 Lund, Sweden.}
\affiliation[b]{Science and Technology Facilities Council, ISIS department, Rutherford Appleton Labs, \\Didcot OX11 0QX, UK.}
\affiliation[c]{European Spallation Source ERIC (ESS), Data Management and Software Centre, Ole Maal{\o}es vej 3, 2200 Copenhagen N, Denmark.}
\affiliation[d]{Laboratory for Neutron Scattering and Imagaing, Paul Scherrer Institut, 5232 Villigen PSI, Switzerland.}
\affiliation[e]{Thin Film Physics Division, Department of Physics, Chemistry and Biology (IFM), Link\"{o}ping University, SE-581 83 Link\"{o}ping, Sweden.}
\affiliation[f]{Division of Nuclear Physics, Lund University, P.O. Box 118, SE-22100 Lund, Sweden.}
\affiliation[g]{Dipartimento di Fisica ``G. Occhialini'', University of Milano-Bicocca, Italy.} 

\emailAdd{francesco.piscitelli@esss.se}

\abstract{The Multi-Blade is a Boron-10-based neutron detector designed for neutron reflectometers and developed for the two instruments (Estia and FREIA) planned for the European Spallation Source in Sweden. A reflectometry demonstrator has been installed at the AMOR reflectometer at the Paul Scherrer Institut (PSI - Switzerland). The setup exploits the Selene guide concept and it can be considered a scaled-down demonstrator of Estia. The results of these tests are discussed. It will be shown how the characteristics of the Multi-Blade detector are features that allow the focusing reflectometry operation mode. Additionally the performance of the Multi-Blade, in terms of rate capability, exceeds current state-of-the-art technology. The improvements with respect to the previous prototypes are also highlighted; from background considerations to the linear and angular uniformity response of the detector.}


\begin{document}
\maketitle
\flushbottom

\section{Introduction}
The Multi-Blade~\cite{MIO_MB2014,MIO_MB2017,MIO_MB16CRISP_jinst,MIO_ScientificMBcrisp,MIO_HERE,MIO_MyThesis} is a Boron-10-based neutron detector designed for neutron reflectometry. This detector technology, originally introduced at the Institut Laue-Langevin (ILL~\cite{ILL,ILL2,ILL3}) and subsequently developed at the European Spallation Source (ESS~\cite{ESS,ESS_TDR,ESS-design,ESS2011}), can be nowadays considered a mature technology ready for deployment at the instruments. 

Neutron reflectometers are a demanding class of instruments in terms of detector requirements. The Multi-Blade is designed for the two ESS reflectometers, Estia~\cite{INSTR_ESTIA,INSTR_ESTIA0,INSTR_ESTIA1,INSTR_ESTIA2} and FREIA~\cite{INSTR_FREIA,INSTR_FREIA2}, for which a sub-millimetre spatial resolution, high signal-to-background ratio, and a high counting rate capability of several kHz per square millimetre~\cite{ESS_TDR,DET_rates,HE3S_kirstein} are required to enable the full power of these instruments.
At present, neutron reflectometers at existing facilities (spallation sources and reactors) are experiencing significant limitations due to saturation occurring at detectors, mainly in terms of rate capability (limited to a few hundreds Hz per square millimetre) and spatial resolution (limited to approximately 1.5 - 2 mm)~\cite{INSTR_D17,INSTR_FIGARO}.

In the past, several techniques have been proposed to improve the operating performance of neutron reflectometers. The methods are based on spin-space~\cite{INSTR_R_Spin2}, time-space~\cite{OTT_tiltof} or energy-space encoding~\cite{OTT_gradtof,OTT_refocus,R_Cubitt1,R_Cubitt2}. In particular the Estia reflectometer at ESS will exploit focusing reflectometry~\cite{INSTR_ESTIA0,OTT_refocus} technique. The method increases total intensity by simultaneously using the neutron wavelength spread and the incident angle divergence with a beam focused beam on the sample. In order to focus the neutrons on the sample a set of two subsequent elliptical neutron guiedes, i.e. the \textit{Selene} guide~\cite{INSTR_ESTIA0,INSTR_ESTIA1}, is used. With this set-up, AMOR~\cite{AMOR_Clemens,AMOR_Gupta,AMOR} at PSI\cite{PSI,PSI_SINQ} can be considered a scaled-down demonstrator of Estia~\cite{INSTR_ESTIA2}. 

In the past years the Multi-Blade technology has been widely characterized and tested at various instruments at different facilities. The proof of concept for this detector technology has been primarily demonstrated with a first test of a prototype at the ILL in 2013~\cite{MIO_MB2014}. A detailed characterization~\cite{MIO_MB2017} of a Multi-Blade detector has been carried out at the Budapest Neutron Centre~\cite{FAC_BNC,FAC_BNC1,FAC_BNC2} in 2015. In 2017 a reflectometry demonstrator has been installed and tested at the neutron reflectometer CRISP~\cite{CRISP1} at the ISIS neutron and muon source in the UK~\cite{ISIS,ISIS1,ISIS2} performing a full detector characterization along with specular and off-specular reflectivity measurements on several samples~\cite{MIO_MB16CRISP_jinst,MIO_ScientificMBcrisp}. Moreover, background studies, in particular the response of this detector to fast neutrons~\cite{MIO_fastn} and gamma-rays~\cite{MIO_MB2017}, have been carried out at the Source Testing Facility (STF~\cite{SF1,SF2}) at the Lund University in Sweden.

In 2018, a Multi-Blade demonstrator has been installed at the AMOR reflectometer at PSI to test the capabilities of the detector in an environment as close as possible to the future Estia instrument. The results of these tests are presented in this manuscript. It will be shown how it is  possible to exploit the advantages of the focusing reflectometry technique thanks to the Multi-Blade technology, particularly due to its spatial resolution and its counting rate capability.

Along with the measurements of high intensity specular reflectivity exploiting the focusing technique, a series of detector characterisation is presented, emphasising how the technical features of the detector improve the scientific output of the experiment. 

\section{Description of the set-up}\label{mbsetup}
\subsection{Multi-Blade Detector}
\begin{figure}[htbp]
\centering
\includegraphics[width=0.99\textwidth,keepaspectratio]{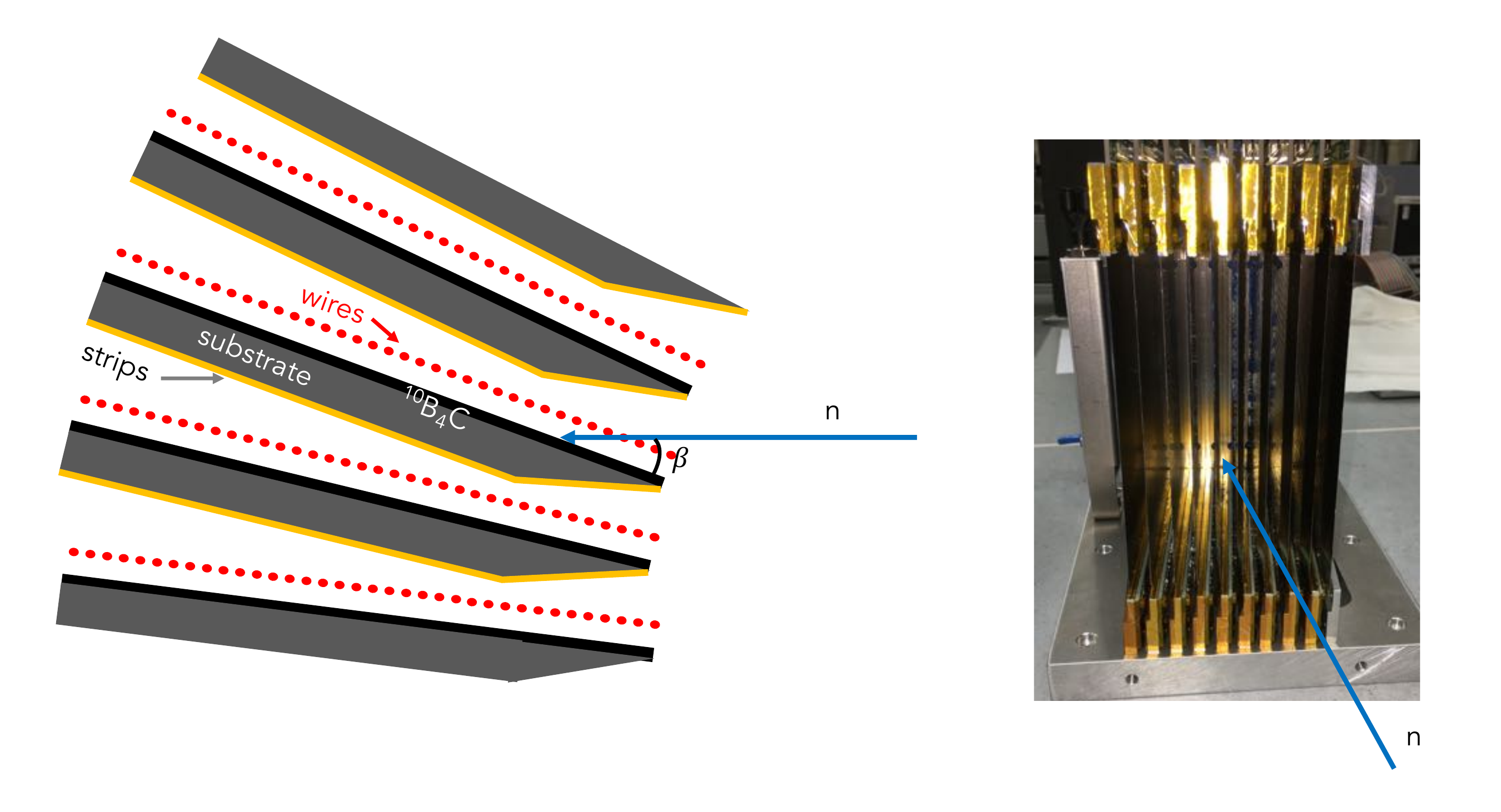}
\caption{\label{mbSketch} \footnotesize Sketch of the cross-section of the Multi-Blade detector composed of identical units (cassettes) arranged over a circle and placed adjacent to each other (left). Each cassette is inclined at an angle $\beta=5^{\circ}$ with respect to the neutron incoming direction. Note that the scale is exaggerated for ease of viewing. Each cassette has a titanium blade coated with a $^{10}$B$_{4}$C layer; the readout is performed through a plane of wires and a plane of strips. A picture of the Multi-Blade detector with Ti-blades made up of 9 units (right).}
\end{figure}

The Multi-Blade is a modular detector made up of several units, here called \textit{cassettes}. A schematic cross-section and a picture of the detector are shown in figure~\ref{mbSketch}. Each cassette is an independent Multi Wire Proportional Chamber (MWPC) flushed with $\mathrm{Ar/CO_2}$ (80/20 mixture) at atmospheric pressure and it is equipped with a two-dimensional readout system; a plane of wires (red dots) orthogonal to a plane of strips (yellow). A $^{10}$B$_{4}$C layer acts as a neutron converter in each unit (black). The coatings were done at the ESS Detector Coatings Workshop in Link{\"o}ping in a DC magnetron sputtering system~\cite{B4C_carina,B4C_carina3,B4C_Schmidt}. The cassettes are arranged over a circle around the sample position, so that each $^{10}$B$_{4}$C layer is inclined at an angle ($\beta$ in  figure~\ref{mbSketch}) of 5$^{\circ}$ with the incoming neutron direction. This has the effect of improving the detection efficiency~\cite{MIO_analyt,MIO_MyThesis} besides the spatial resolution across the wires and the counting rate capability. 

Each cassette has 32 wires and 32 strips. A 32-channel FET-based charge pre-amplifier board is connected to each plane of wires and strips. Therefore the channels are readout individually. Each 32-channel board is connected to a CAEN V1740D digitiser (12 bit, 62.5 MS/s)~\cite{EL_CAEN}. Six digitisers are available in total, each one can readout 64 channels, i.e. one cassette. Hence, six cassettes can be readout simultaneously during the tests. The digitisers are synchronized and share the same clock; i.e. the time-stamp which is associated to any event is common across the digitisers. Time-stamping allows for Time-of-Flight (ToF) measurements. This system is fully asynchronous; any channel above a threshold (set in the hardware) is recorded independently. A DPP-QDC (Digital Pulse Processing) firmware was used, thus only a value (QDC) proportional to the energy released on a wire (or a strip) is recorded and not the full signal trace. Note that the signals are shaped, hence any value among amplitude, pulse integral (QDC) or time-over-threshold (ToT) gives the same information: a value related to the energy released on a wire or a strip that can be used to compute the Pulse-Height-Spectrum (PHS). The raw data contains the channel number and its time-stamp. Note that a single neutron event generally triggers a group of channels: multiplicity is generally larger than one~\cite{MIO_MB16CRISP_jinst}. A MATLAB~\cite{MATLAB:R2019a_u3}-based software is used on the raw data to identify clusters of groups of channels which denotes a single neutron event. A further threshold is applied in the software to each channel in order to reduce background, e.g. by rejecting events caused by gamma radiation. The energy released in the gas volume by a neutron conversion fragment ($\alpha$ or $Li$ particles) is then reconstructed by summing all energies (QDC) belonging to the same cluster. This electronics and software reconstruction is the same used for the tests at the CRISP reflectometer, more details can be found in~\cite{MIO_MB16CRISP_jinst}. Note that the software threshold is applied to all the plots shown in this manuscript according to the considerations that will be given in section~\ref{parax_unif}.

Every cluster, or neutron event, is then identified by its spatial coordinates $(X,Y)$ and a time $(T)$, i.e. its Time-of-Flight. Moreover, a single event has an associated energy (QDC) released on all wires (or strips) involved which can be used to calculate the PHS. The energy information can be used to reject background by properly selecting a software threshold and, as it will be shown in section~\ref{parax_unif}, to compensate the wire to wire gas gain variation due to the detector geometry. 

Note that the spatial coordinates $(X,Y)$ reflect the physical channels in the detector: 32 strips and 32 wires respectively. They represent the projection over the detector entrance window ( i.e. the projection of the cassette's planes toward the neutron incoming direction). The ToF $(T)$ of each event is the time of arrival of a neutron to the specific wire ($Y$ on the vertical axis). When $(T)$ has to be translated to neutron wavelength ($\lambda$), the physical depth ($Z$ along the neutron incoming direction) of each wire must be taken into account~\cite{MIO_MB16CRISP_jinst}. 

\subsection{Experimental set-up on AMOR}\label{amorsetup}
A Multi-Blade detector has been installed at the AMOR reflectometer~\cite{AMOR_Clemens,AMOR_Gupta,AMOR} at SINQ/PSI ~\cite{PSI,PSI_SINQ}. AMOR is a versatile instrument which allows measurements in Time-of-Flight (ToF) mode and in monochromatic ($\theta-2\theta$) mode using either polarized or unpolarized neutrons. 

The measurements have been performed in the ToF mode, with a wavelength accessible range $2.5< \lambda < 15$ \AA. The spectral distribution of wavelengths is peaked around 4 \AA. A cold neutron beam hits a double chopper system, which defines the starting signal for the Time-of-Flight calculation of the neutrons reaching the detector. The chopper discs have two phase coupled openings of 13.6$^{\circ}$ and the chopper-detector distance can be varied between 3.5 and 10 m. The first disk opening and second disc closing phases are equal, which allows a constant relative wavelength resolution ($\sigma_{\lambda}/\lambda$)~\cite{doublediscchopper}. At AMOR, with the used detector distance, this corresponds to $\sigma_{\lambda}/\lambda \approx 2\%$~\cite{INSTR_ESTIA2}. 

\begin{figure}[htbp]
\centering
\includegraphics[width=1\textwidth,keepaspectratio]{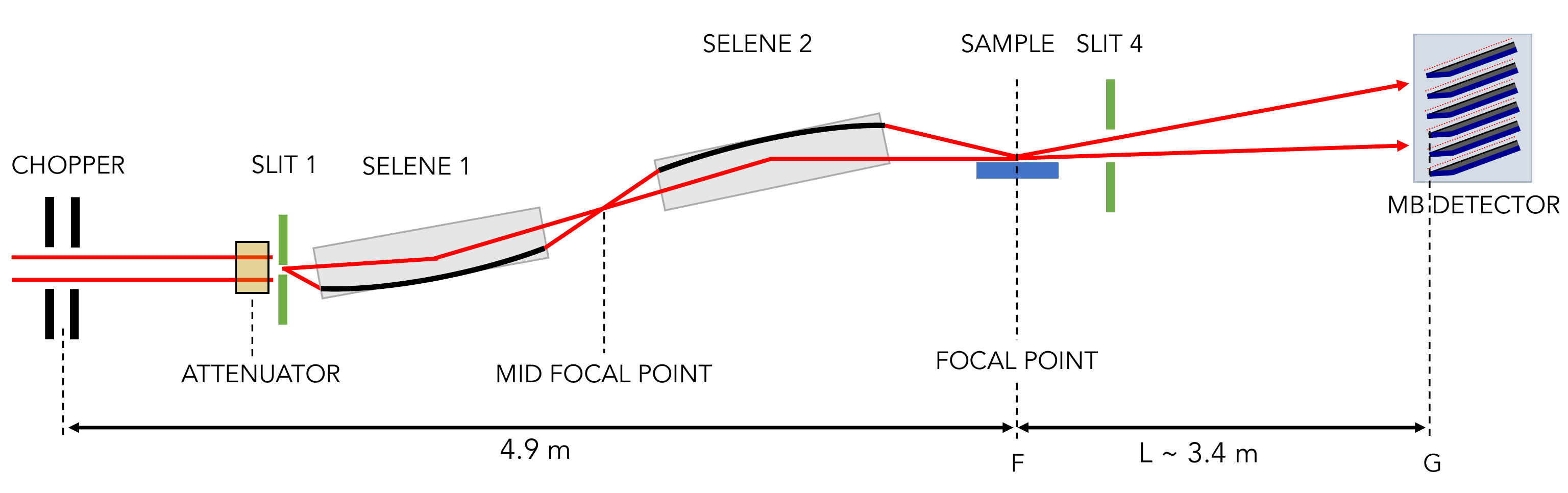}
\caption{\label{amor} \footnotesize Schematic view of the AMOR reflectometer at PSI.}
\end{figure}

Each of these parameters can be selected independently, therefore the resolution of the instrument can be adjusted in the range $\Delta q /q=1-10\%$. In figure~\ref{amor} a sketch of the instrument layout is shown.  

The key feature of this instrument is the \textit{Selene} guide~\cite{INSTR_ESTIA0,INSTR_ESTIA1,INSTR_ESTIA2}, which allows to use the high-intensity specular reflectometry~\cite{OTT_refocus,R_Cubitt2,OTT_general} with a broad converging beam focused at the sample. The subsequent divergent beam can be detected by a position-sensitive detector; a sub-millimetre spatial resolution is required to have a good $\theta$-resolution at small angles. Therefore, the requirements of high spatial resolution and high counting rate are mandatory for the detector operation. The same concept will be adopted for the Estia reflectometer~\cite{INSTR_ESTIA} at ESS.

Note that AMOR is a horizontal reflectometer (vertical scattering plane), therefore the detector is oriented with a high vertical resolution, as depicted in figure~\ref{amor}. The $Y$ coordinate refers to the plane of wires, while the $X$ to the plane of strips. This notation is used for all the results discussed in the following sections.   

\subsection{DMSC integration} \label{sec:DMSC}
For the first time, the Multi-Blade data was acquired using the ESS software data acquisition system (provided by the DMSC\footnote{DMSC: Data Management \& Software Centre}) in a configuration similar to the one that will be used when ESS is in operation~\cite{DMSC_Brigh,essghal}. The data path consists of a temporary readout system that collects readouts from commercial CAEN digitisers~\cite{EL_CAEN}.  These readouts are then transmitted as UDP packets over Ethernet to the Event Formation Unit (EFU)~\cite{DMSC_EFU}, which is a modular C++ application running on standard linux-based PCs. The EFU performs clustering and coincidence processing and calculate the time and position (actually the detector pixel) of the incident neutrons. The (time, pixel) tuples define the events, which are then streamed to a data aggregator based on Apache Kafka~\cite{DMSC_kafka}. In addition to the streaming of events, the EFU saves the raw readouts to HDF5 files~\cite{DMSC_HDF} as reference for later analysis. From the aggregator it is possible to create live detector images and other visualizations using a custom tool named Daquiri~\cite{DMSC_daquiri}. For a detailed description of the software based event processing at ESS see~\cite{MB-DMSC}.

\section{Results with high-intensity Estia-like mode and the Multi-Blade detector}\label{sec:estia}

The similar experimental conditions between AMOR and Estia give the unique possibility to perform measurements similar to the future operation of the reflectometer at ESS. The Multi-Blade detector is the chosen technology to be employed for this class of instrument at ESS, due to the distinctive features tuned for neutron reflectometry applications~\cite{MIO_MB16CRISP_jinst,MIO_ScientificMBcrisp}. The combined measurement using an Estia-like instrument set-up and the Multi-Blade, allows to prove the detector operation in a high-intensity specular reflectometry~\cite{INSTR_ESTIA} environment as similar as the one proposed by Estia~\cite{INSTR_ESTIA1}.

The reflectivity of a reference sample has been measured following the procedure developed on AMOR~\cite{INSTR_ESTIA2}. The simulation of a Ni/Ti multilayer sample is described in the Estia proposal~\cite{INSTR_ESTIA} and it is used to show the principles of the high-intensity specular reflectometry measurements. Therefore, a $10 \times 10\,$mm$^2$ Ni/Ti multilayer was used to perform the reflectivity measurements with the Multi-Blade detector at AMOR.

In the high-intensity operation mode the full divergence and the full wavelength range supplied by the neutron guide are used. The wavelength resolution is given by the pulse length of the source, because $\lambda$ is encoded in the Time of Flight (ToF)~\cite{INSTR_ESTIA2} as described in section~\ref{amorsetup}. The angular resolution ($\sigma_{\theta}$) is determined by the detector spatial resolution (FWHM $\Delta Y \approx 0.5\,$mm) and the sample-to-detector distance (L$\approx 3.4\,$m), which gives $\sigma_{\theta} \approx 0.004^{\circ}$.

\begin{figure}[htbp]
\centering
\includegraphics[width=0.49\textwidth,keepaspectratio]{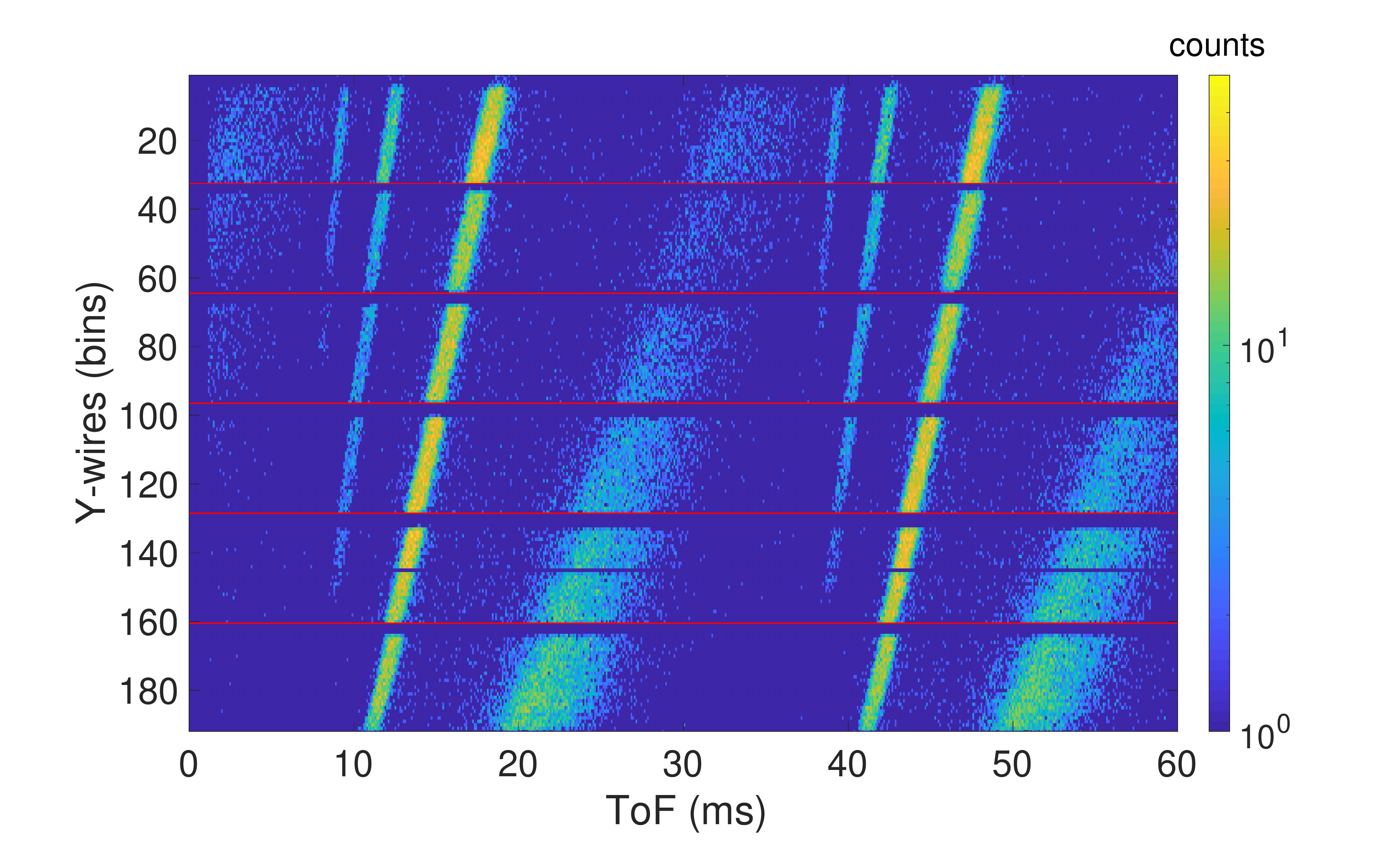}
\includegraphics[width=0.49\textwidth,keepaspectratio]{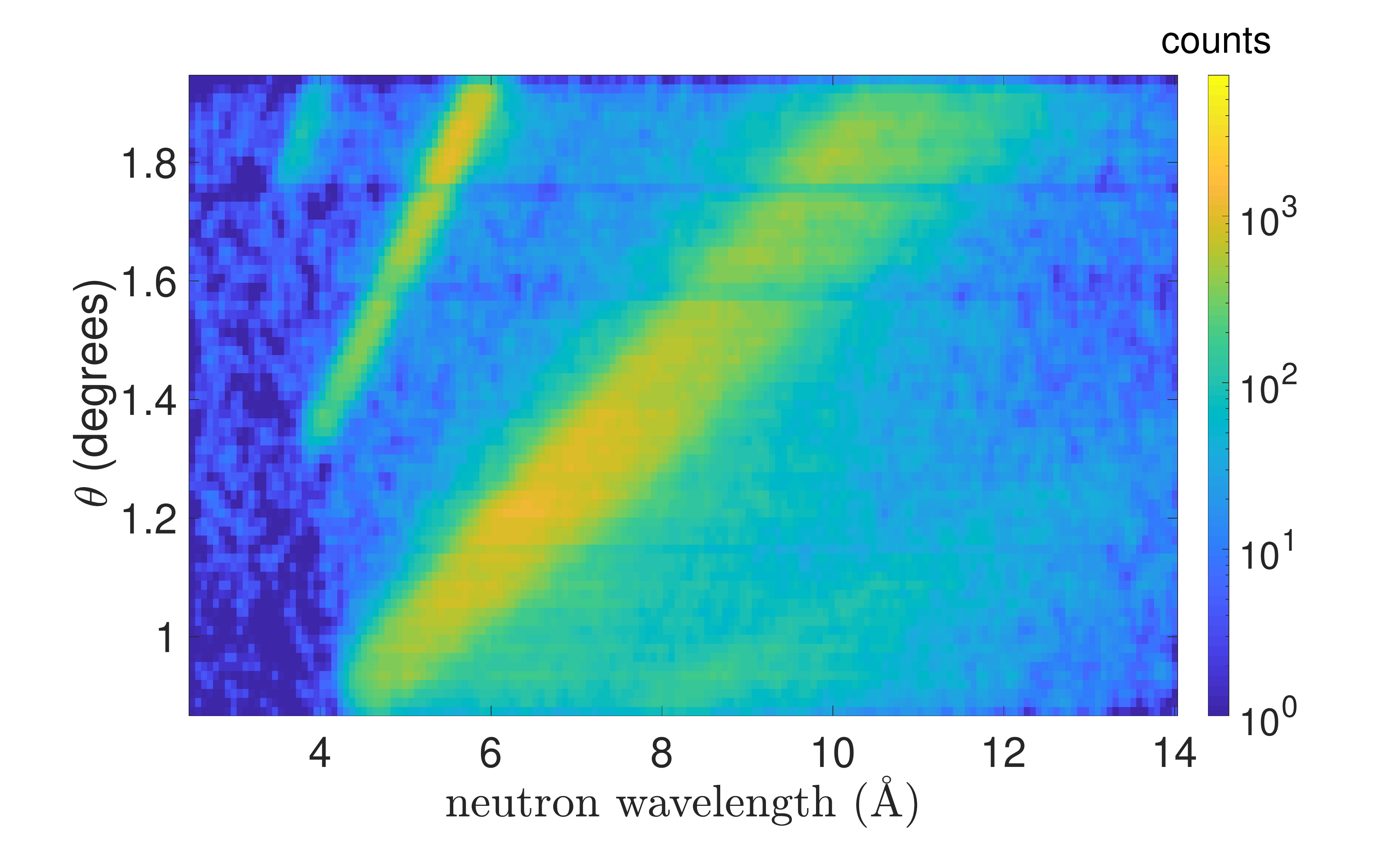}
\caption{\label{nititof} \footnotesize Measured reflectivity intensities $I(T,Y)$ (left) and $I(\lambda,\theta)$ (right) for a Ni/Ti multilayer sample at an angle $\omega = 2.2^{\circ}$. The color scale represents counts in a whole acquisition run.}
\end{figure}

As mentioned in section~\ref{mbsetup}, each detected neutron can be parameterized with $T$, $X$ and $Y$ for the Time-of-Flight and the position on the detector. The neutron wavelength can be calculated from the instrument parameters and $T$. The intensity $I$ in the 2-D space $I(T,Y)$, depending on the Time-of-Flight and on the spatial coordinate, $Y$ is considered in the case of a vertical scattering plane reflectometer. The $I(T,Y)$ can be subsequently transformed into a $I(\lambda,\theta)$ map; the two corresponding maps are shown in figure~\ref{nititof}, respectively. The double-disk chopper opens twice for every reset of the ToF and the two bunches of neutrons are visible in the $I(T,Y)$ map. The horizontal red lines are drawn to identify the six cassettes of the detector. At each line, few channels (2-3) do not show any counts. Note that these are not dead channels, it is instead the shadowing effect inherent to the detector geometry, a detailed description of which can be found in~\cite{MIO_MB16CRISP_jinst}. Physically the last firing wire of one cassette is adjacent to the first wire of the neighbour cassette. Each shadowed wire can be removed from any further analysis, without losing actual data.

\begin{figure}[htbp]
\centering
\includegraphics[width=0.49\textwidth,keepaspectratio]{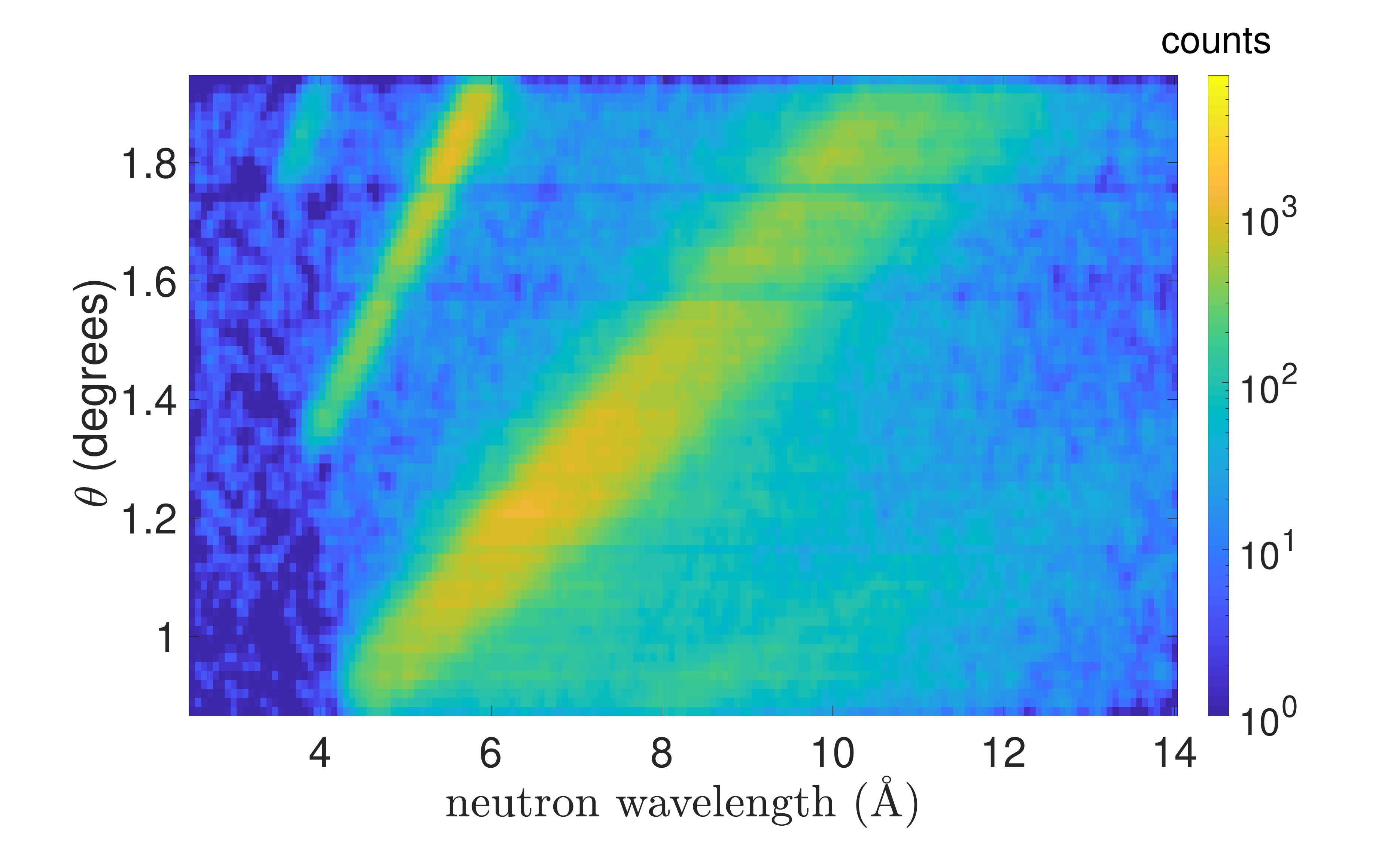}
\includegraphics[width=0.49\textwidth,keepaspectratio]{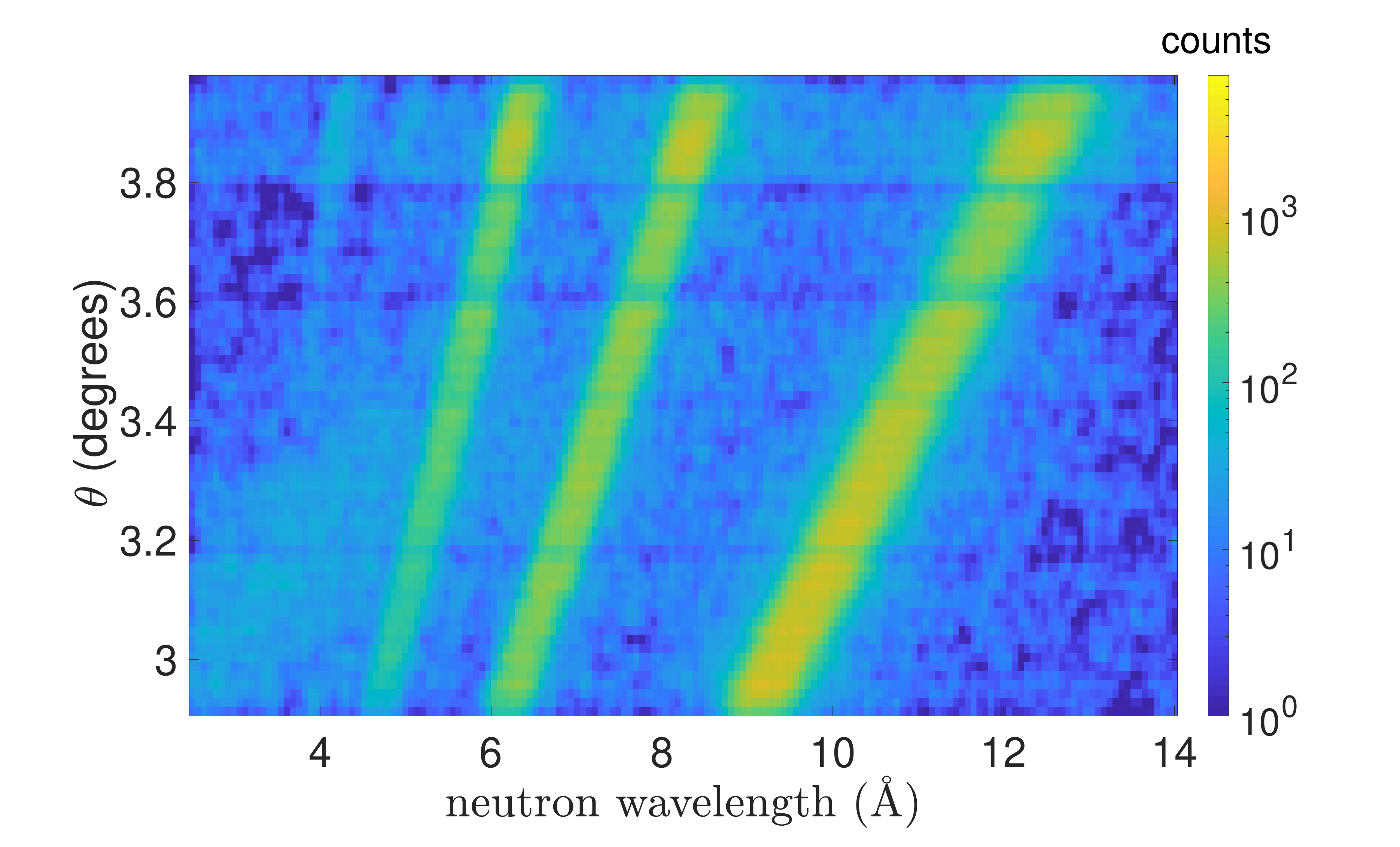}
\caption{\label{nitiangles} \footnotesize Measured reflectivity intensities $I(\lambda,\theta)$ for $\omega = 1.4^{\circ}$ (left) and $\omega = 3.2^{\circ}$ (right). The color scale represents counts in a whole acquisition run.}
\end{figure}

Figure~\ref{nititof} shows on the right the intensity once it is transformed in the $\lambda-\theta$ space (the two bunches of neutrons in the ToF are recombined in a single bunch in neutron wavelength). Each row corresponds to a ToF measurement at a given $\theta$ and each column corresponds to an angle dispersive measurement for a given $\lambda$. The shadowed wires have been removed, a region of lower intensity is normally observed because of the charge collection at the first wire of each cassette. The decrease in efficiency in this region is never worse than 50\% and occurs in one wire, a detailed discussion can be found in~\cite{MIO_MB16CRISP_jinst}. A small intrinsic non-uniformity due to the modular design of the detector can be observed. Indeed, the response of each cassette slightly varies. The difference between the cassettes can be observed in figure~\ref{nititof} and~\ref{nitiangles}. A reduced intensity is recorded in the top part of the active area. The reasons for the non-uniformity will be discussed in detail in section~\ref{parax_unif}.

The $q_z$ range needed for the measurement determines the $\omega$ and $\theta$ settings. Here $\omega$ is the angle between the incoming beam center and the sample, while the detector is centered at $\theta= 2\omega$. For most applications at the AMOR  reflectometer, angles between $1^{\circ} \lesssim \omega \lesssim 4^{\circ}$, are sufficient to cover a wide $q_z$-range, approximately [0.004,0.23] \AA$^{-1}$~\cite{INSTR_ESTIA2}.

Three angles have been measured: $\omega = 1.4^{\circ}$, $\omega = 2.2^{\circ}$ and $\omega = 3.2^{\circ}$. The recorded intensity is shown in figure~\ref{nititof} in phase space $(\lambda-\theta)$ for $\omega = 2.2^{\circ}$, while figure~\ref{nitiangles} $I(\lambda,\theta)$ depicts it combined for the remaining two measured angles.

\begin{figure}[htbp]
\centering
\begin{tabular}{cc}
\includegraphics[width=0.49\textwidth,keepaspectratio]{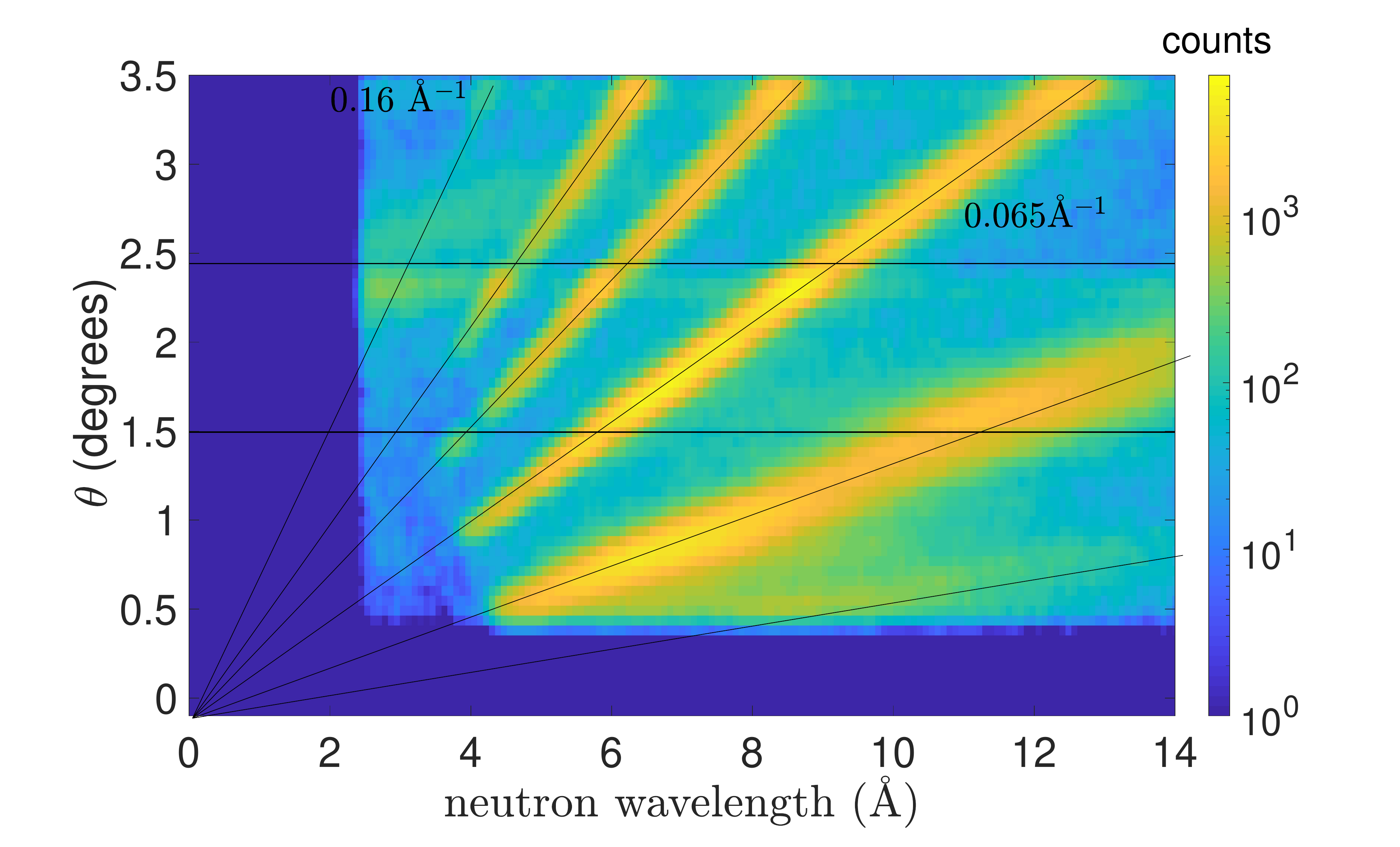}&
\includegraphics[width=0.45\textwidth,keepaspectratio]{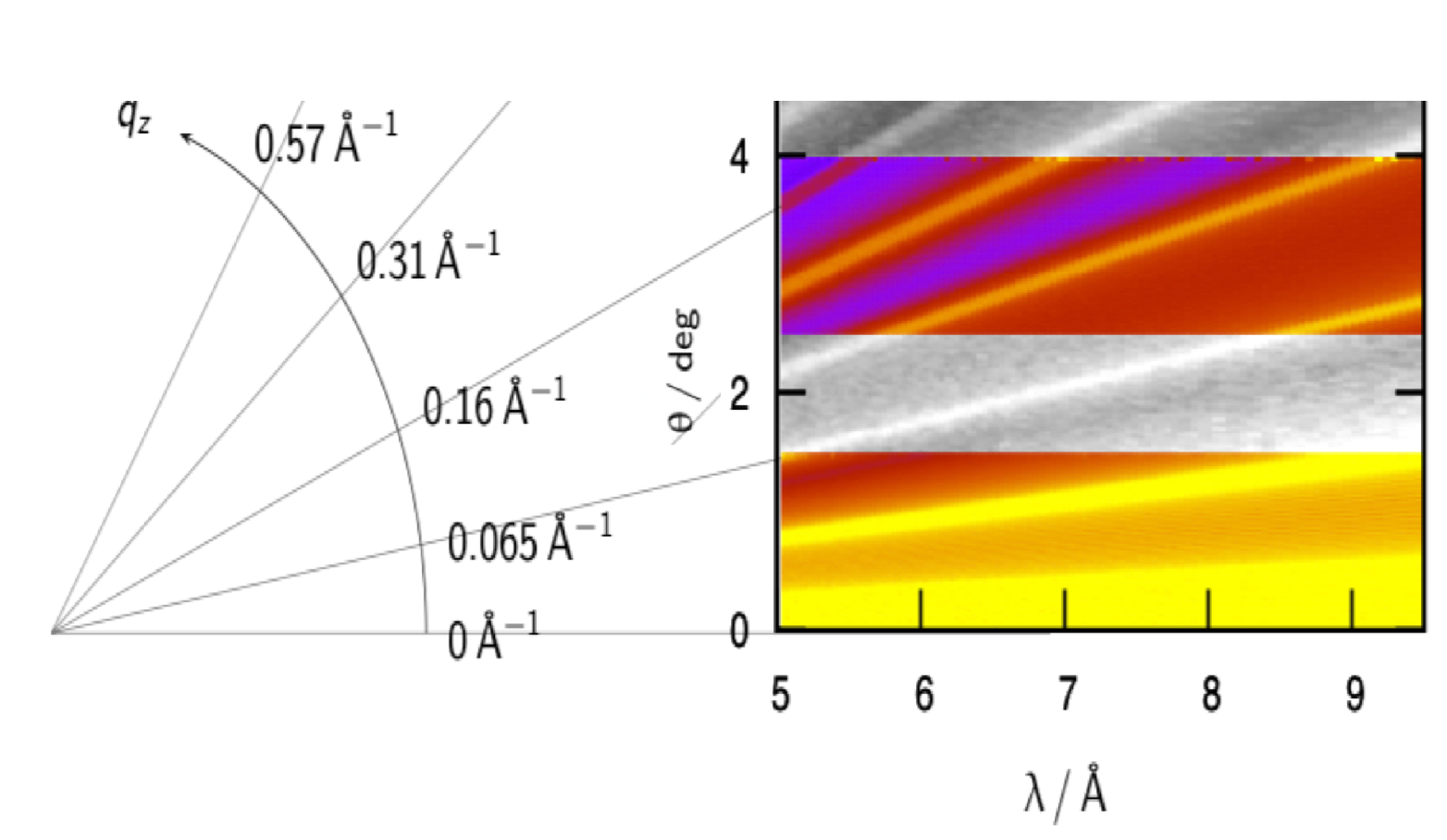} \\
(a) & (b) \\
\end{tabular}
\caption{\label{nititot} \footnotesize (a) Combination of the three measured angles of a Ni/Ti multilayer sample in the $I(\lambda,\theta)$ space, the two horizontal black lines indicate the overlap of two adjacent datasets. The color scale represents counts in a whole acquisition run. (b) Simulation of the same sample for high intensity reflectivity measurement. From the Estia instrument proposal~\cite{INSTR_ESTIA}.}
\end{figure}

The combination of the three measurements is shown in figure~\ref{nititot} (a), where the critical edge and five Bragg peaks of the Ni/Ti sample are visible. The horizontal lines identify the three regions measured by each angle. A high reflectivity, for neutron wavelengths that satisfy the Bragg condition, is expected for a multilayer composed of equal thickness bilayers. The first five order Bragg reflections have been measured with $\omega <4^{\circ}$ as shown in figure~\ref{nititot} (a). A similar behaviour is obtained with the simulations described in~\cite{INSTR_ESTIA} and shown in figure~\ref{nititot} (b). The five Bragg peaks in the simulations are represented with the yellow lines and the corresponding $q_z$ is shown, e.g. the second peak for $\lambda = 5$ \AA{}, is at $\theta \approx 1.5^{\circ}$ or $q_z  = 0.065\,$\AA$^{-1}$, while the fifth peak has a $q_z  = 0.16\,$\AA$^{-1}$. Note that, the location of the Bragg-peaks is not significant, since the simulation does not necessarily have the same bilayer periodicity of the measured sample.

\begin{figure}[htbp]
\centering
\includegraphics[width=0.6\textwidth,keepaspectratio]{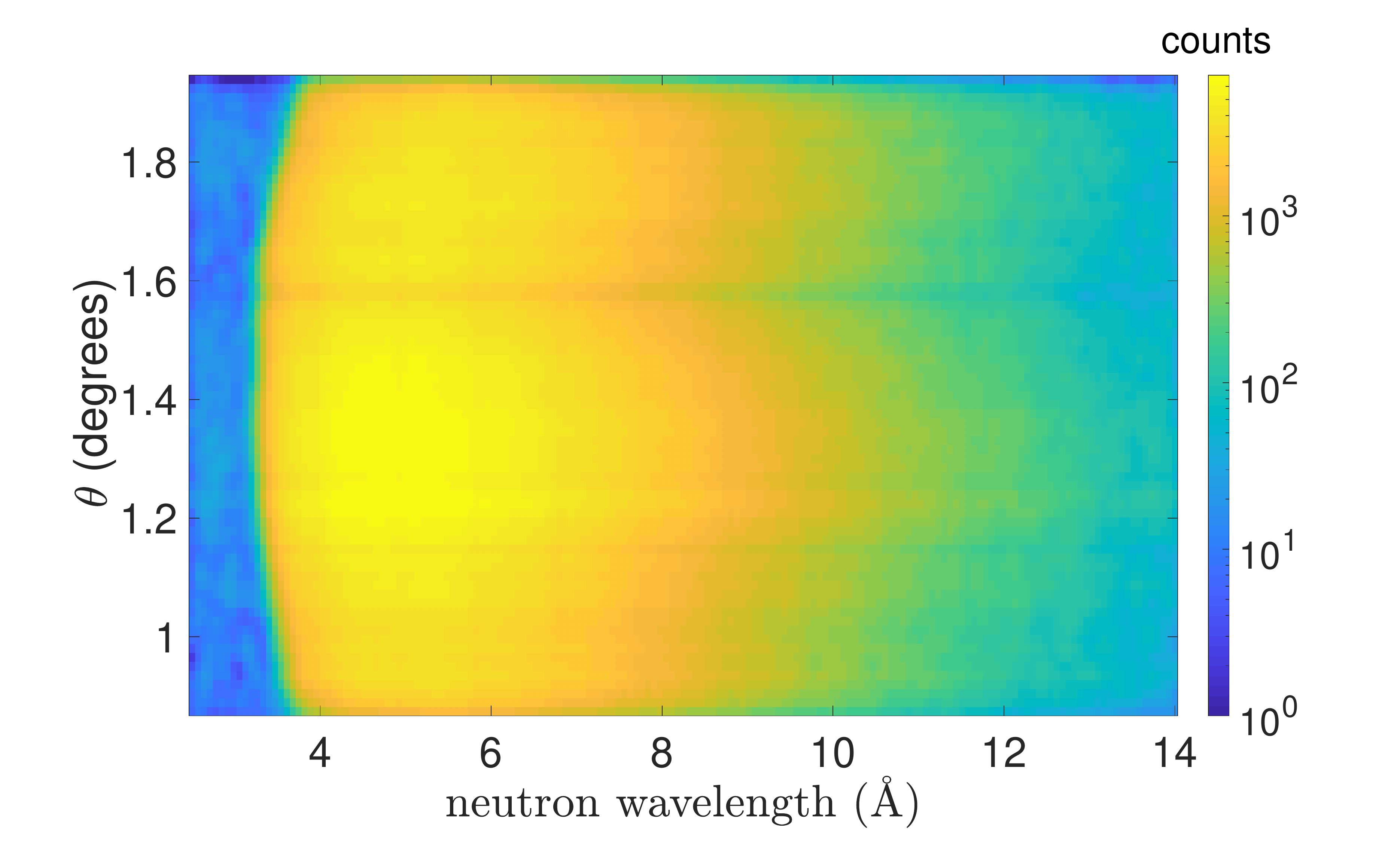}
\caption{\label{sm} \footnotesize Intensity map in the $\lambda-\theta$ space for a $m=5$ Ni/Ti supermirror. This measurements is used to normalise the reflectivity measurements. The color scale represents counts in a whole acquisition run.}
\end{figure}

In order to obtain the specular reflectivity, $R(q_z)$, the intensity must be normalized to a known reference sample with a high reflectivity over a wide $q_z$-range, measured under similar condition. The sample used is a $m=5$ Ni/Ti supermirror with a known reflectivity $R(q_z)$ and same dimensions of the Ni/Ti multilayer sample ($10 \times 10\,$mm$^2$), in order to ensure a comparable footprint. This was measured at $\omega =1.4^{\circ}$ and the intensity profile is shown in figure~\ref{sm}. The uniform response of the supermirror over a wide $\lambda$ range is obtained as expected. The inhomogeneities can be corrected, because a pixel-by-pixel normalisation can be applied on the reflected intensities dividing it by the reference measurement intensity (figure~\ref{sm}). Note that the less intense horizontal lines originate from the imperfections of the Selene guide prototype. A detailed explanation of the data reduction method applied can be found in~\cite{INSTR_ESTIA2}. 

The measurements in the high-intensity mode of a known sample indicate good results both on the operation of the Multi-Blade detector and on the effectiveness of this technique for neutron reflectometry experiments. The requirements concerning the features of the detector are particularly demanding to measure under these conditions. Hence, a set of measurements were performed to investigate the detector capabilities. The counting rate capability, the study of the background and the detector uniformity are among the most crucial to accomplish high quality reflectometry experiments, together with the spatial resolution beyond the present limitations. The results are presented in the following section.

\section{Detector performance}\label{meas}

\subsection{Counting rate capability}\label{rate}
\begin{figure}[htbp]
\centering
\includegraphics[width=0.6\textwidth,keepaspectratio]{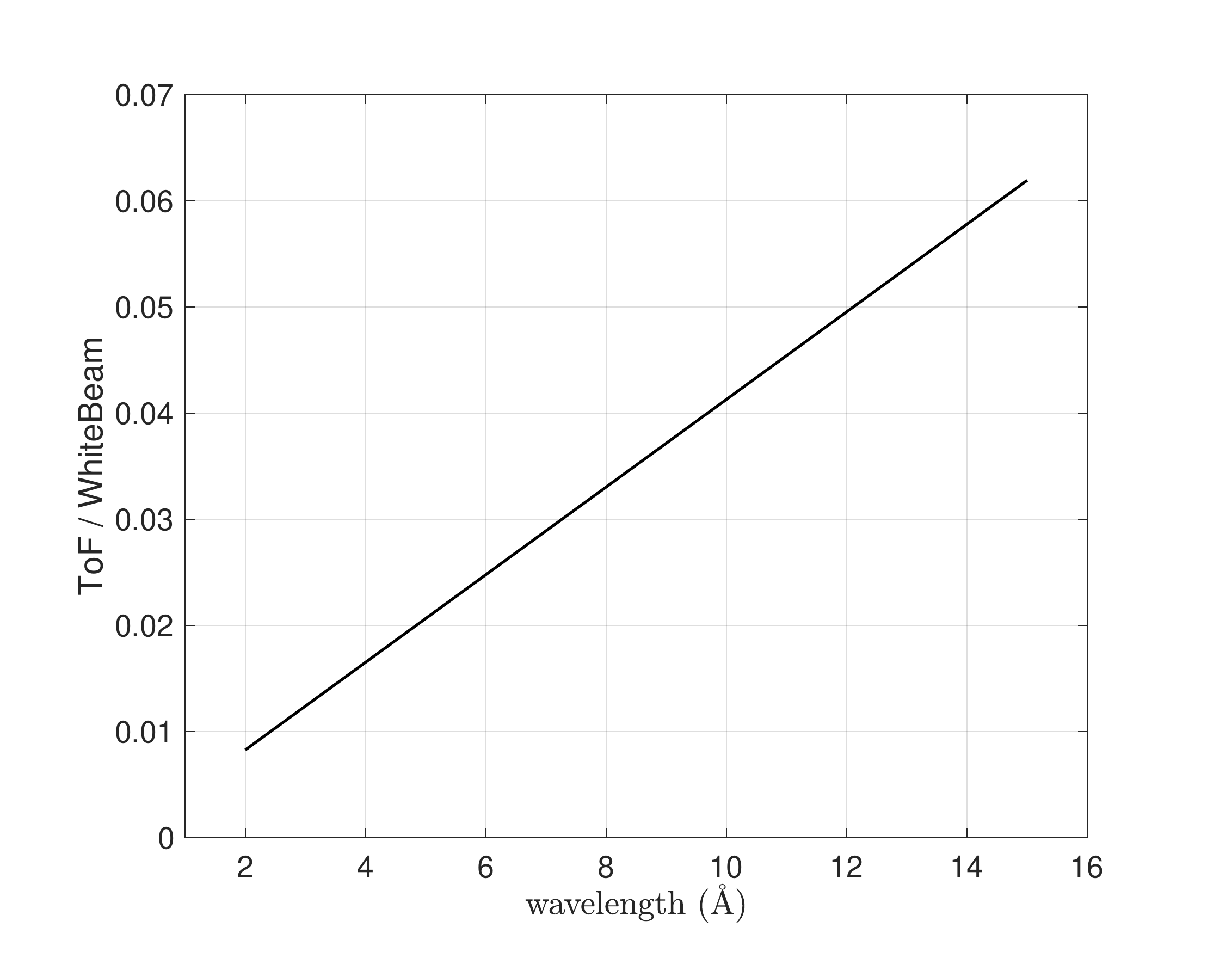}
\caption{\label{chop} \footnotesize Double-blind chopper configuration at 1000~rpm (two openings), the intensity factor depends on the wavelength linearly.}
\end{figure}

The counting rate capability is one of the most demanding requirement for detectors for neutron reflectometry. A set of measurements were performed to investigate the Multi-Blade response to a high rate environment. The flexibility of the AMOR instrument allows to move the detector along the beam line, between a maximum sample-to-detector distance of 4\,m and the focal point determined by the {\it Selene} guide, i.e. the sample position. The measurements of the direct beam were performed at two fixed positions: $G$ with sample-to-detector distance $L \approx 3.4\,$m and $F$ at the focal point; these reference points are depicted in figure~\ref{amor}. 
\\ The global time-averaged rate can be expressed as the total number of counts per second recorded in the total active area of the detector. The local instantaneous peak rate is defined as the highest instantaneous neutron count rate on the brightest detector pixel~\cite{DET_rates}. 
\\ The full divergence of the beam, given by the geometry of the {\it Selene} guide, is $\Delta \theta=1.6^{\circ}$. At the position $G$ the beam covers an area $A \approx 95 \times 95\,$mm$^2$, whereas the full detector active area is $A_D \approx 130 \times 60\,$mm$^2$. At the focal point ($F$) the full beam intensity is focused on a few mm$^2$.
\\The measurements at the two locations allow to compare a maximum local count rate with the same rate spread over a wider detector area. Therefore, eventual loss in counting rate can be understood to come from detector limitations. 
\\The measurement of the local rate in position $F$ was performed with a continuous white beam, i.e. the chopper system was stopped, leading to an increasing time-averaged neutron flux of about a factor 50. Due to the double-blind chopper configuration, the intensity factor depends on the wavelength linearly between 3\AA\, and 15\AA. The ratio between the white beam and the ToF mode is approximately 80 at 3\AA \, and 16 at 15\AA\, (see figure~\ref{chop}).

The global rate measurements in location $G$ was instead performed in ToF mode with the chopper system spinning. Note that the measurement of the global rate was performed in a condition of moderate rate to ensure that no saturation of the detector occurs and as a benchmark for the local rate measurement as it will shown in this section. 
\\With a continuous beam, the intensity is constant over time and it can be considered an instantaneous quantity when divided by the duration of the measurement. In this case the detector experiences the most extreme conditions because the space charge in the gas volume accumulates and no recovery time is foreseen for the detector. In contrast, the Time-of-Flight mode when the beam peak intensity is instead, of the order of a few $\mu$s and, even if  the detector cannot cope with the rate at peak, the subsequent moderate rate allows the charge to be evacuated and the electric field to be restored.
\\Several measurements have been performed, in the two experimental set-up positions $G$ and $F$, attenuating the beam with attenuators, steel slabs $(h=90\times l=28\times w=10)\,$mm$^3$. These attenuators have been placed in the mid focal point of the \textit{Selene} guide (see figure~\ref{amor}) across the beam. Three measurements have been carried out: (A) when no attenuation is used, (B) one slab was employed and (C) with the beam attenuated by two slabs. 
\\The attenuation factor given by each attenuator was measured with the detector in position $G$ operating the chopper at a 16\,Hz rotation frequency. As mentioned before, in this configuration the detector works far from any saturation regime, therefore it will be used as reference measurement for the attenuation factors. The ratio between the integrated number of counts of (A), (B) and (C) gives the attenuation factors: 
\begin{align}
	& f_{AB}=\frac{\int I_{(A)}}{\int I_{(B)}} = 4.9 \pm 0.3 \label{AttFact1} \\
	& f_{AC}=\frac{\int I_{(A)}}{\int I_{(C)}} = 21.2 \pm 1.2 \label{AttFact2}
\end{align}

\begin{figure}[htbp]
\centering
\includegraphics[width=0.7\textwidth,keepaspectratio]{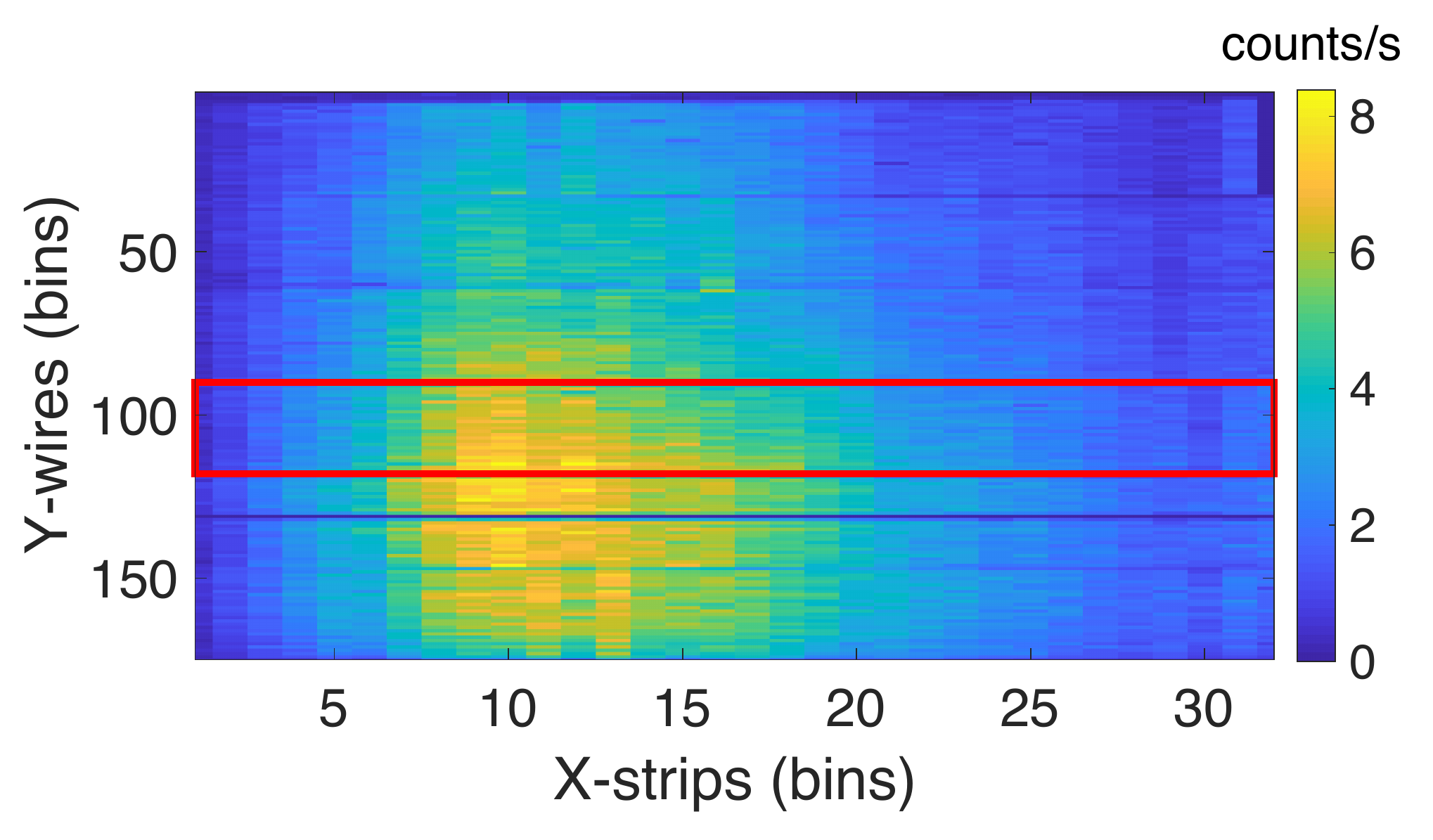}
\caption{\label{Rate2Dg} \footnotesize 2D image of the six cassettes of the Multi-Blade placed in position G, sample-to-detector distance 3.4 m; divergent beam on the detector for the global rate measurement. The red mark highlights the beam direction of the local rate measurements. A bin in the vertical scale (wires) is 0.35\,mm and in the horizontal scale (strips) is 4\,mm.}
\end{figure}

The uncertainties have been calculated through the error propagation considering a statistical fluctuation of counts.
\\The rate comparison between the global and local measurements can be verified by both the intensity factor due to the double-blind chopper and the attenuation factors of equations~\ref{AttFact1} and~\ref{AttFact2}. It is, therefore, possible to distinguish whether the detector has endured the local rate or not.   
\\Figure~\ref{Rate2Dg} shows the global rate measured in location $G$, configuration (A) (no attenuation). The counts are normalized by time in the detector active area $A_D$. The global time-averaged rate is approximately 17~kHz, a similar rate is recorded with the $\mathrm{^3He}$ detector employed at the instrument without saturation effects. This value is in agreement with the expected rate in this instrument configuration. In average each detector pixel records several counts per second, see figure~\ref{Rate2Dg}, the operation regime for both the Multi-Blade and the He-3-based AMOR detector is below any saturation either due to electronics limitations or physical detector limitations. Indeed, the actual limits of the Helium-3 technology is around hundreds Hz/mm$^2$~\cite{INSTR_FIGARO}, and the Multi-Blade detector is expected to achieve the order of kHz/mm$^2$~\cite{MIO_MB2017}. Therefore, in this configuration, the total rate is recorded by the detector and it can be used for the comparison with the local instantaneous peak rate measurement. 

\begin{figure}[htbp]
\centering
\includegraphics[width=0.9\textwidth,keepaspectratio]{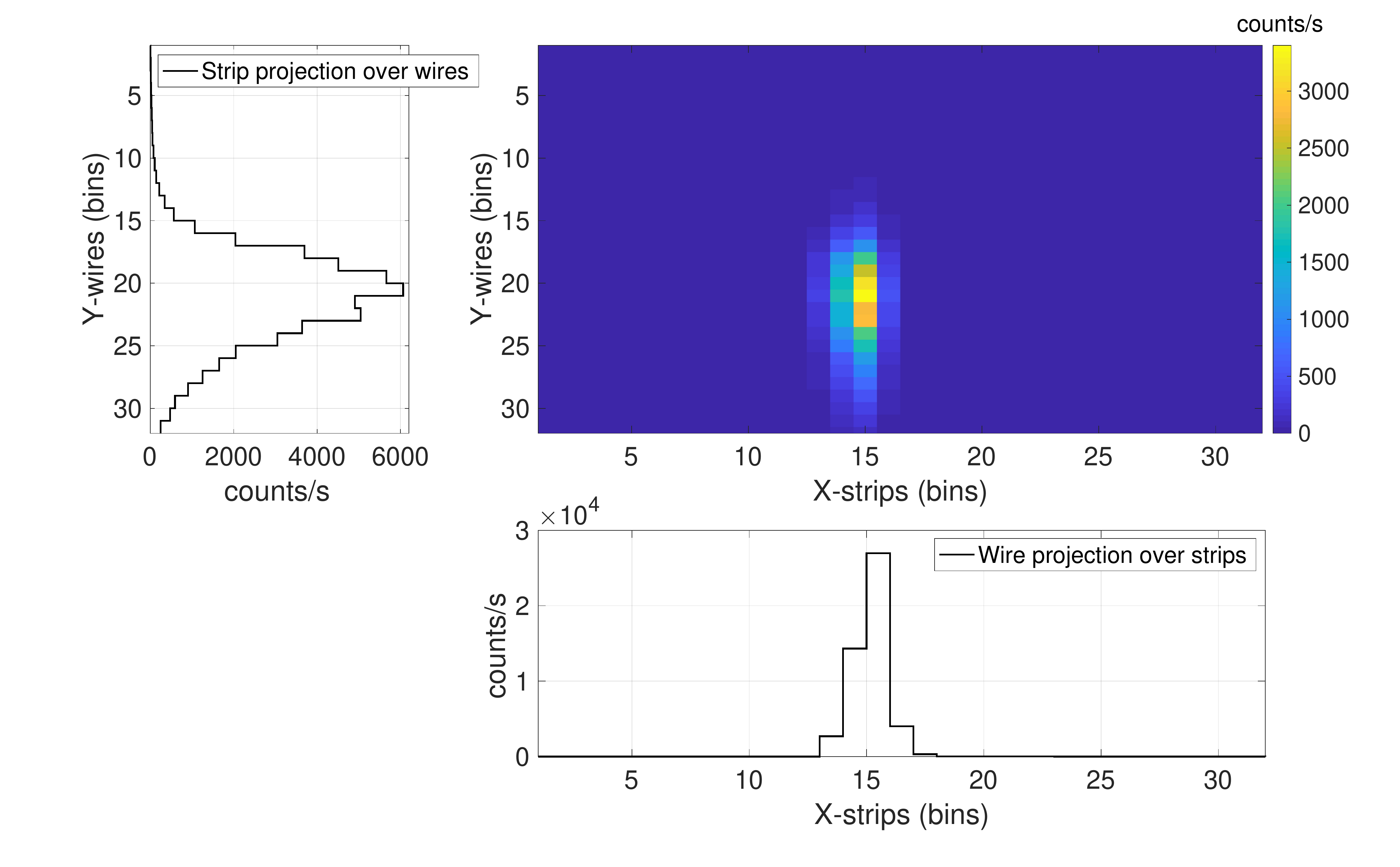}
\caption{\label{Rate2Dl} \footnotesize 2D image for the measurement of the local rate when the detector is placed in the position F at the focal point. The beam extends no more than in one cassette, which is marked with the red rectangle in figure~\ref{Rate2Dg}. The projections along the two dimensions $(X)$, strips summed over wires, and $(Y)$ wires summed over strips is shown. A bin in the vertical scale (wires) is 0.35\,mm and in the horizontal scale (strips) is 4\,mm.}
\end{figure}

The dark wire in figure~\ref{Rate2Dg} is a noisy channel, whose intensity has been reduced by applying a high software threshold (see section~\ref{parax_unif} for more detail), while the strip channel no. 32 of the top cassette is used to readout the monitor of the instrument for normalization purpose. The red rectangle highlights the cassette illuminated for the measurement performed in position $F$. 
\\In neutron reflectometry the most challenging scenario from the detector point of view regards the local rate; for most applications a high rate ($>$ kHz) is centred in a small area ($<$ mm$^2$).
The intensity in the 2-D spatial coordinate $I(X,Y)$ is depicted in figure~\ref{Rate2Dl}. A bin in the vertical scale (wires) corresponds to 0.35\,mm and in the horizontal scale (strips) to 4\,mm.
\\This measurement has been performed in configuration (C), with two attenuation slabs, in location $F$. The largest recorded intensity per unit time (i.e. under a constant irradiation over time), is $I(X,Y) \approx 3.4\,$kHz in a detector pixel. A pixel is about 1.4 \,mm$^2$ ($0.35 \times 4$mm$^2$). The dead-time due to pile-up events can be calculated as shown in~\cite{DET_knoll} and it depends mainly on the type of front-end electronics used. The probability for two, or more, pile-up events for such a rate is below $2\%$. i.e. a correction factor of 1.017 can be applied for this rate to correct for pile-up.
\\Figure~\ref{Rate2Dl} shows, as well, the projections of both strips $(X)$ and wires $(Y)$ over the other coordinate, respectively. The incoming beam spreads over an area of approximately $(7\times 16)\,$mm$^2 = 112\,$mm$^2$. The two projections show the intensity of the beam profile for each wire over the all firing strips (figure~\ref{Rate2Dl}, left), and for each strip over all firing wires (figure~\ref{Rate2Dl}, bottom).

The rate in 112 mm$^2$ is approximately 50\,kHz. It can be noticed that the adjacent pixels, with respect to the brightest one (Y-wire = 21, X-strip = 15), have less intensity due to the beam distribution, but they can in principle tolerate the same rate as the brightest pixel. The rate recorded along the brightest strip is approximately 27\,kHz for 20 firing wires (area $\approx 28\,$mm$^2$), while the intensity of the brightest wire, summing over the illuminated strip is about 6 kHz (area $\approx 5.5\,$mm$^2$).
Therefore, the rate capability does not scale linearly with the area of the detector. A larger value than 50\,kHz in 112 mm$^2$, which corresponds to a twelfth of a cassette, can be in principle recorded. Each detector module consists in 1024 pixels, considering the measured lower limit of few kHz per pixel, the counting rate of one cassette can be on the order of MHz. This value matches the present read-out electrics limitation, which is theoretically 880\,kHz per digitizer~\cite{EL_CAEN} (note that one cassette is readout by one digitizer).   
\\One can conclude that a local rate of the order of a few kHz in $\approx$1\,mm$^2$ is the measured rate at which the Multi-Blade detector does not show any saturation. Note that, this is the maximum rate that has been measured experimentally and the real rate capability of the detector, given with the standard definition of 10\% loss due to the dead time, is somewhere above this value. This result is in accordance with the requirements set by the Estia instrument, which has been estimated 7\,kHz$\cdot$mm$^{-1}$ at the full ESS power operation of 5\,MW~\cite{ESS_TDR}. 

\begin{figure}[htbp]
\centering
\begin{tabular}{ccc}
	& {\bf ABSOLUTE RATE} & {\bf NORMALIZED RATE} \\
	\rotatebox{90}{\hspace{2.5cm} {\bf 	GLOBAL RATE}}   
	& \includegraphics[width=.44\textwidth,keepaspectratio]{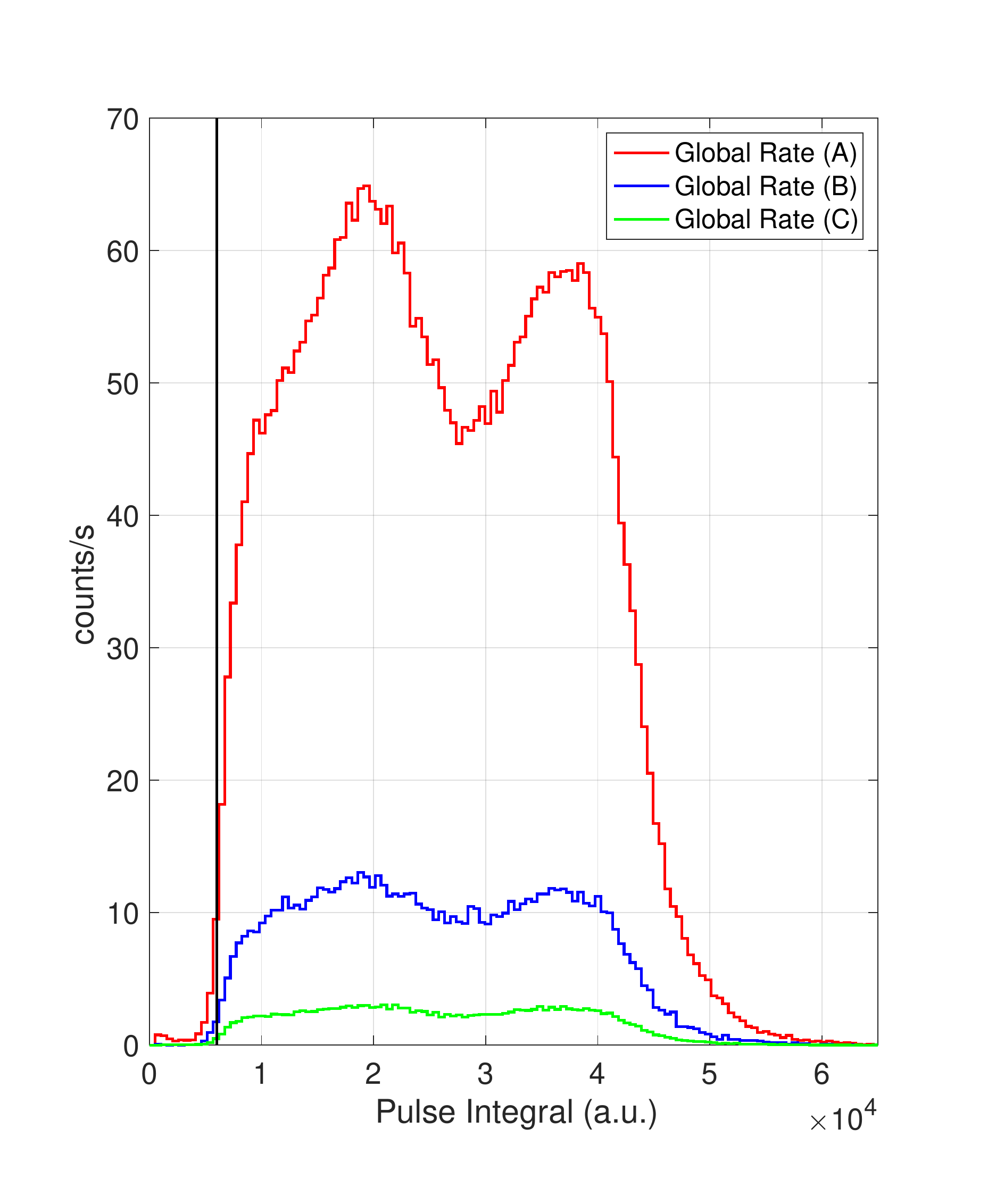}& 
	    \includegraphics[width=.44\textwidth,keepaspectratio]{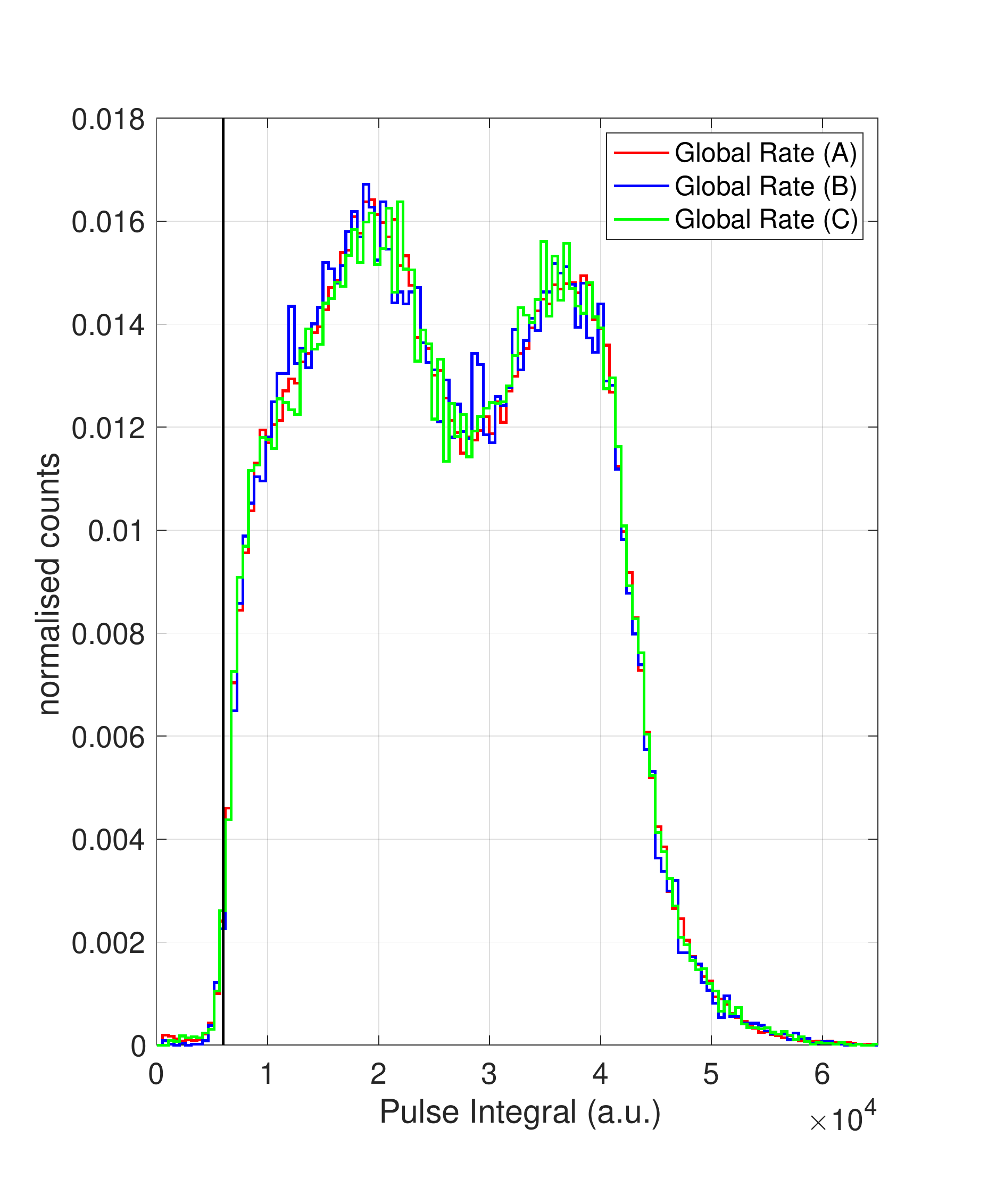} \\    
	    & (a) & (b) \\ 
     \rotatebox{90}{\hspace{2.5cm} {\bf LOCAL RATE}}   
     & \includegraphics[width=.44\textwidth,keepaspectratio]{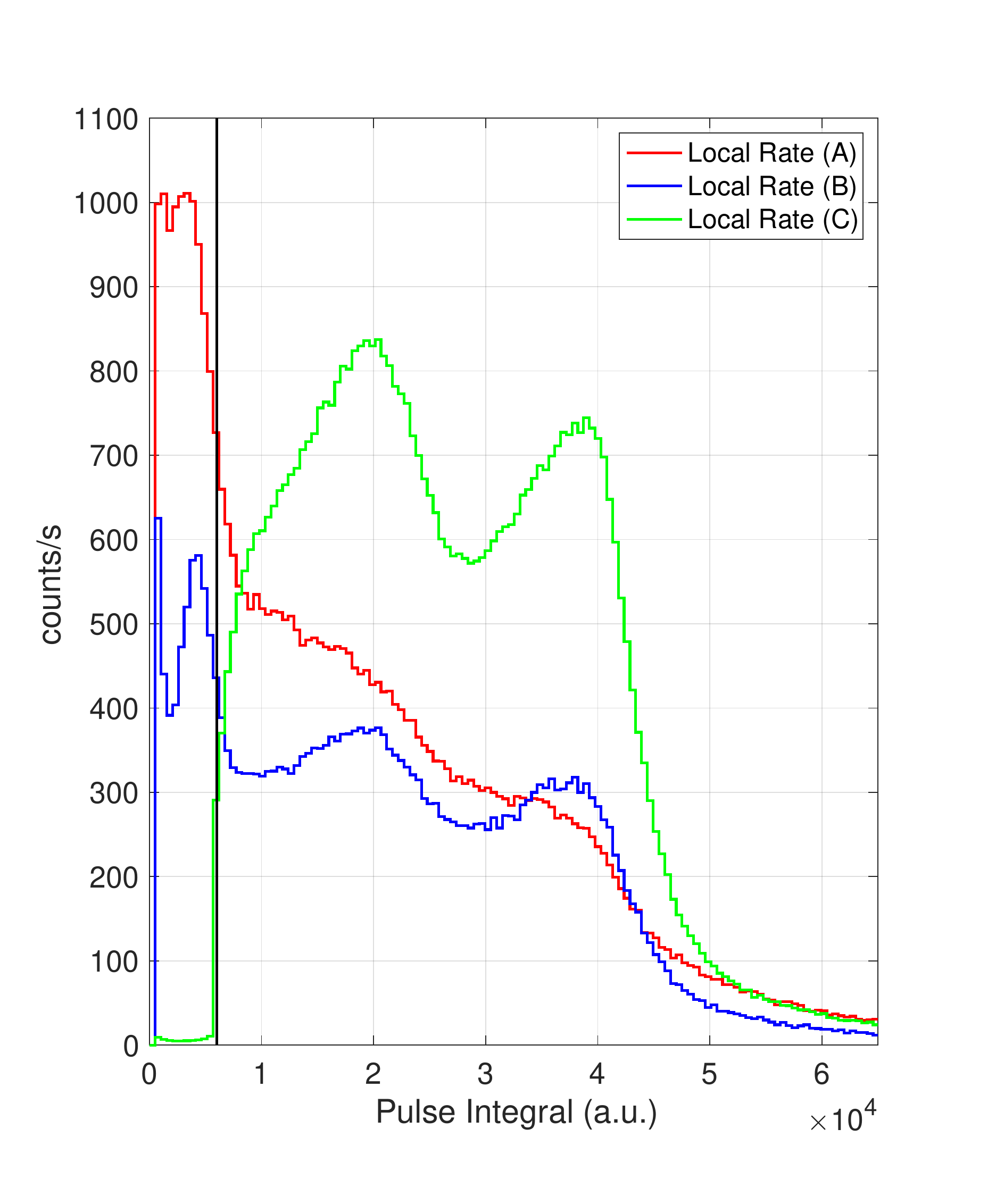}& 
        \includegraphics[width=.44\textwidth,keepaspectratio]{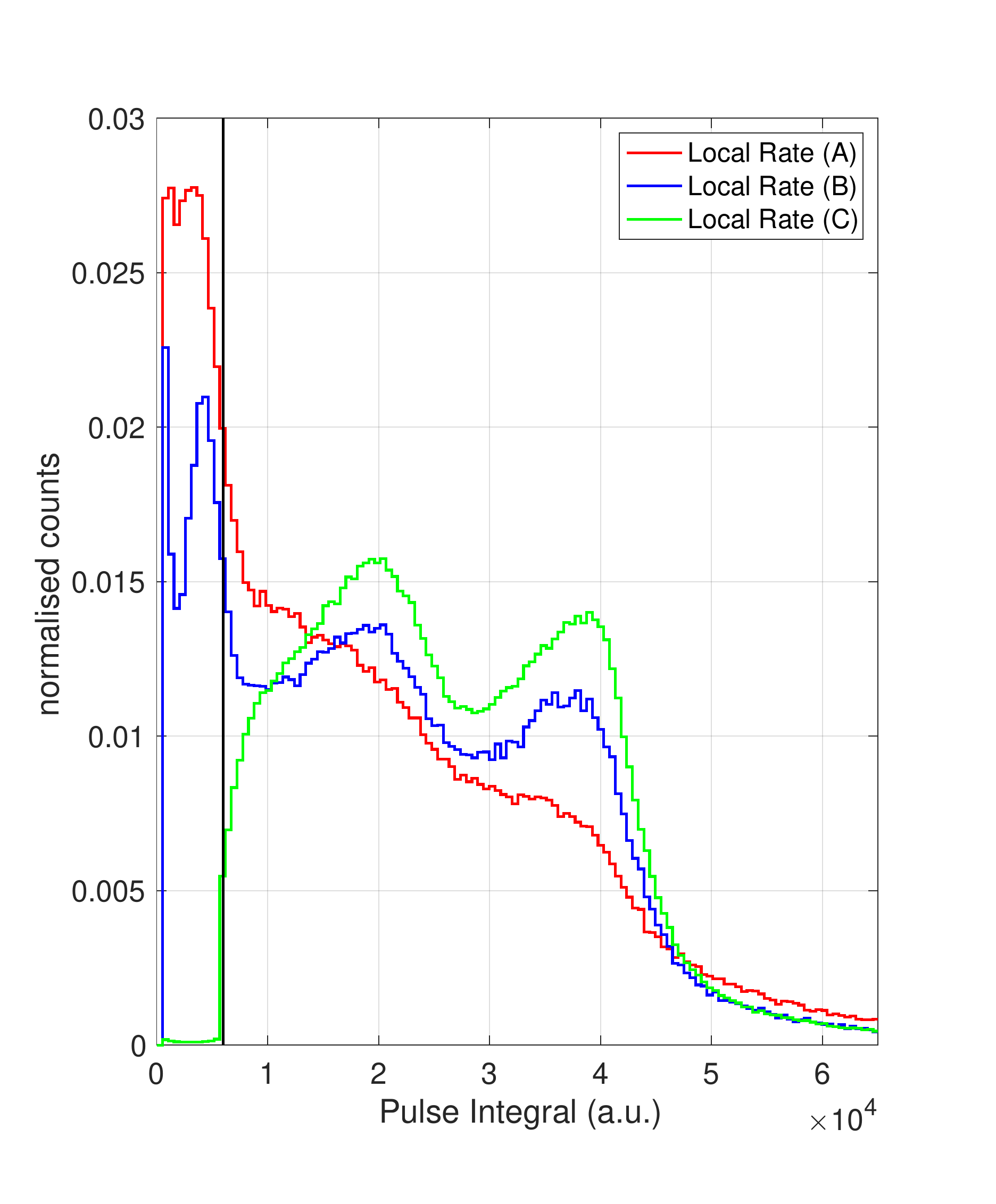}\\ 
      & (c) & (d) 
	\end{tabular}
\caption{\label{Ratephs} \footnotesize (a) PHS for the measurement of the global rate with the detector at the position G in three conditions: (A) when no attenuation is used, (B) one attenuator and (C) the beam was attenuated by two attenuators. (b) same PHS as (a) normalised by their integral. (c) PHS for the measurement of the local rate with the detector at the position F and the three attenuation factors. (d) same PHS as (c) normalised by their integral. The vertical black line corresponds to approximately a threshold of 200\,keV.}
\end{figure}

As previously mentioned, the ToF to white beam mode intensity factor is approximately 50. In order to compare the two rate measurements, configuration (A) for the global ($\approx$17 kHz) and (C) for the local ($\approx$50 kHz) rate, the attenuation factor $f_2$ (equation~\ref{AttFact2}) has to be considered. A factor of about 60 is obtained. Since the area covered by the beam $A$ is larger than the detector active area $A_D$ this value cannot be used as a precise method of discrimination. Nevertheless, the two values are of the same order of magnitude and it can give a first insight of eventual event losses. A quantitative investigation is performed based on the study of PHS.
\\As already mentioned in section~\ref{mbsetup}, each event is recorded with a certain energy (QDC or Pulse Integral, on the horizontal axis in the plots in figure~\ref{Ratephs}), therefore the study of the variation of pulse height distribution gives important information regarding the charge generated by the radiation interaction and the inherent response of the detector. The PHS recorded for neutron conversion in Boron has a characteristic shape defined by two peaks, which correspond to the energy released by the yields of the interaction, i.e. $\alpha$ or $Li$ particles~\cite{MIO_analyt,MIO_MB16CRISP_jinst}, background events, e.g. gamma-rays, result in a sharp peak at low energies. 

The PHS for the three measurements (A), (B) and (C) at the fixed position $G$ is shown in (a) of figure~\ref{Ratephs}. The pulse height normalized by the integral of the counts are depicted in figure~\ref{Ratephs} (b), the shape is similarly reproduced for all measurements. No gain shift can be observed in the PHS for any attenuation factor. The PHS are shown here without a software threshold applied in order to visualize the events, that in case of saturation, are shifted toward smaller amplitudes and consequently below this threshold. The black vertical line in the plots represents the software threshold that would have been applied to this data to reject background events, and corresponds approximately to 200\,keV.

The attenuator factors $f_{AB}$ and $f_{AC}$ calculated in equations~\ref{AttFact1} and~\ref{AttFact2}, can be compared with the ratio between the PHS normalisation factors $N_{f_{AB}}= \frac{\int PHS_{(A)}}{\int PHS_{(B)}} $ and $N_{f_{AC}} = \frac{\int PHS_{(A)}}{\int PHS_{(C)}}$. One can define the ratio $\Delta_{x}$ as $\Delta_{x}=\frac{f_x}{N_{f_x}}$ which defines the agreement between the expected attenuation in the measurements. The more $\Delta$ differs from 1 the larger the count loss. The values are summarized in table~\ref{tabfact}. 

\begin{table}[htbp]
	\centering
	\caption{\label{tabfact} \footnotesize Ratio between PHS normalisation Factors, $N_{f_{AB}}$ and $N_{f_{AC}}$, and $\Delta$ ratio between attenuator factors, $f_{AB}$, $f_{AC}$ and $N_{f_{AB}}$, $N_{f_{AC}}$. The calculation is performed for both set of measurements of global and local rate. Errors on local rate are not given since the detector is saturated and not all events recorded.}
	\smallskip
	\begin{tabular}{|l|c|c|c|c|}
	\hline
	\hline
	 & \bfseries $N_{f_{AB}}$ & \bfseries $N_{f_{AC}}$ & \bfseries $\Delta_{AB}$ & \bfseries $\Delta_{AC}$\\ 
	\hline
	Global Rate Measurements  & $5.1\pm 0.3$ & $21.2\pm 1.9$ & $0.97\pm 0.11$ & $1.00\pm 0.15$ \\
	 \hline
	Local Rate Measurements  & $1.314\pm 0.015$ & $0.685\pm 0.007$ & $3.7 \pm 0.3$ & $31 \pm 2$\\
	 \hline
	 \hline
	\end{tabular}
\end{table}

The global rate of the direct beam at the detector is lower than 20\,kHz, at these rates no saturation is expected at the detector nor at the electronics, as confirmed by the analysis of the PHS. The agreement between the normalization and the attenuation factors $\Delta_{AB}=0.97\pm 0.011$ and $\Delta_{AC} = 1.00\pm 0.15$ is a further proof that the incoming neutron beam is fully recorded.

The PHS obtained by measuring the local rate at the focal point $F$ with a continuous beam for the direct beam (A) and the two attenuated measurements (B) and (C) are shown in (c) of figure~\ref{Ratephs}, the PHS normalised by the integral of the counts are shown in (d). Note that these measurements are performed under extreme conditions, indeed the full continuous beam intensity is focused onto a few pixels. Only when the beam is attenuated with two attenuators (C), the PHS shape is reproduced with no gain shift. In both configurations (A) and (B) a loss is observed. 
\\The PHS shifts towards the lower amplitudes, i.e. below the software threshold. The detected energy depends on the spatial distribution of the charge released by the ionised particles, the higher the rate, the higher the space charge effect that can significantly reduce the amplitude of the original pulse. This effect is visible in figure~\ref{Ratephs} (c) and (d) for (A) and (B), the peak at lower amplitudes is an artefact due to the misleading energy detection of a reduced signal compared to its initial charge. The analysis of the PHS without applying the software thresholds underlines this argument. Indeed, either for the the global rate measurements in $G$, configuration (A), (B) and (C), and for the local rate measured in $F$, configuration (C), the events below threshold are negligible, as expected when the coincidence selection between wires and strips signals is performed~\cite{MIO_MB16CRISP_jinst}. The peaks below threshold observed in (A) and (B), figure~\ref{Ratephs} (d), arise from the events that suffer the energy decrease due to the strong local space charge effect induced by the high incoming rate. 
As shown in table~\ref{tabfact}, the normalisation factors differ by approximately 4 for the measurements with one attenuator (B), and by about a factor 30 without beam attenuation (A), respectively. 
The study of PHS ensure that the local instantaneous rate can be derived from the measurements (C), a lower limit of approximately 3.4\,kHz in a detector pixel can be detected with the Multi-Blade with no saturation and with two or more pile-up events below 2\%. Typically, the limitation of a detector the rate capability is given with a 10\% loss. 
In this campaign of measurements, configuration (B) has an expected rate about 5 times larger than the one recorded in (C). The large amount of data loss, approximately 75\%, is partly caused also by the electronics limitations. 
\\The local instantaneous peak rate 3.4\,kHz is the maximum measured rate without saturation of the Multi-Blade detector. It is already more than an order of magnitude (i.e. a factor $\approx 20$) higher than the state-of-the-art detector technologies used in neutron reflectometry~\cite{INSTR_D17,INSTR_FIGARO}. 

\subsection{Scattering in the detector: effect on the background}\label{bkg}
The Multi-Blade detector is a stack of identical units. Each unit contains a titanium blade, which is coated with a $^{10}$B$_{4}$C layer and acts as a neutron converter. It has been discussed in~\cite{MIO_MB2017} that above a 3 $\mu$m thickness of the coating, the efficiency of the detector is saturated. Any extra layer thickness above 3 $\mu$m only serves as a neutron absorber. The minimum calculated $^{10}$B$_{4}$C layer thickness which ensures an absorption probability above $99\%$ at 2.5~\AA\ is 7.5 $\mu$m. As discussed in~\cite{MIO_MB16CRISP_jinst}, it is desirable to have this extra material in the detector. Neutrons that do not contribute to the signal are absorbed in the layer and are prevented from being scattered by the titanium substrate. Consequently they cannot be detected in other areas of the detector resulting in spurious events, i.e. background. The detector tested at the CRISP reflectometer at ISIS had blades coated with only 4.4 $\mu$m of $^{10}$B$_{4}$C~\cite{MIO_MB16CRISP_jinst}. This thickness was, indeed, not sufficient to keep the background, generated by the neutrons that are not absorbed in the coating and scattered by the blade underneath, within the requirements set by the instruments. For this reason, the present detector tested at AMOR has been equipped with coatings above 7.5 $\mu$m thickness. 

The effects of the scattering in the Multi-Blade detector have been simulated using GEANT4 ~\cite{g1} and the results are presented in~\cite{MIO_MBscattsimu}; the simulation results have been compared with the data taken at CRISP showing a good agreement. Here, a further comparison between measurements at CRISP and AMOR is presented. In section 3.1 of~\cite{MIO_MBscattsimu} the definition of scattered events within the detector is given. To summarise, a neutron can be either scattered on the detector window and detected away from the initial incoming direction or scattered by the blades because it has not been stopped efficiently by the converter layer. 
 
\begin{figure}[htbp]
	\centering
	\includegraphics[width=0.7\textwidth,keepaspectratio]{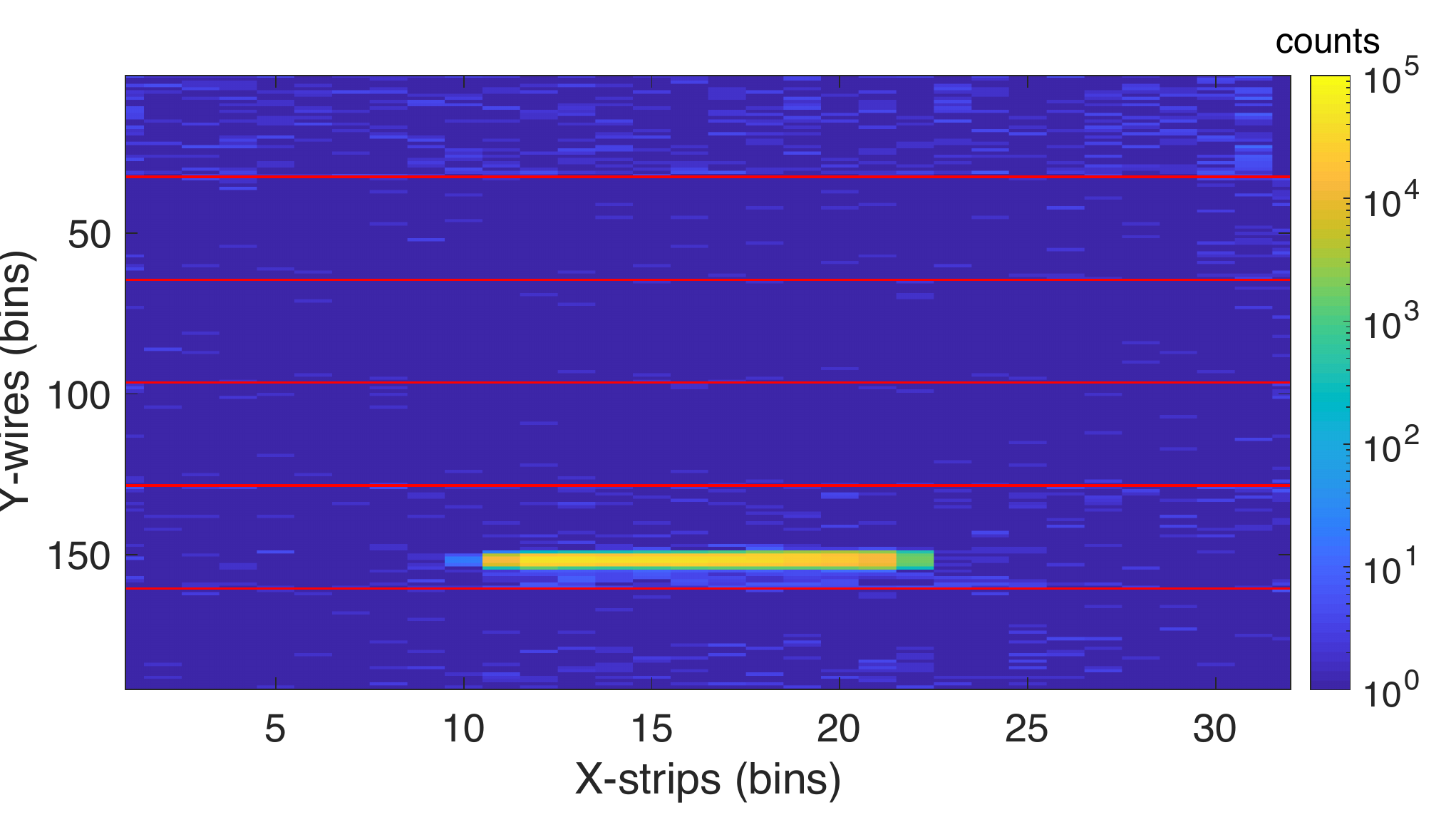}
	\includegraphics[width=0.7\textwidth,keepaspectratio]{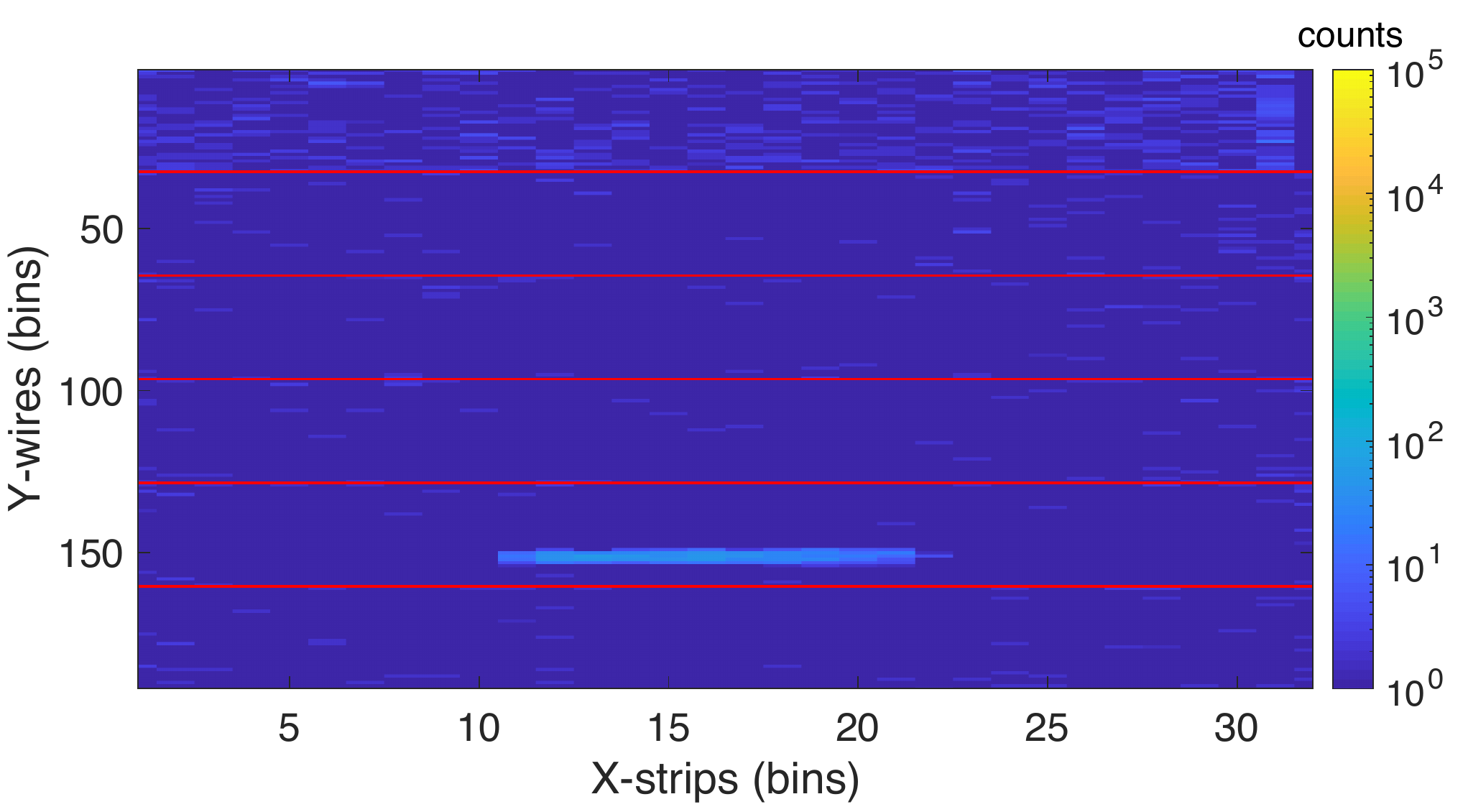}
	\caption{\label{2Dbkg} \footnotesize 2D image of six cassettes of the Multi-Blade detector. Beam transmitted (top), beam blocked at the sample position through an absorber (bottom). The color scale represents counts in a whole acquisition run.}
\end{figure}

In order to investigate the effect of the scattering, the direct beam was directed on the second cassette from the bottom of the Multi-Blade, see figure~\ref{amor}. The beam was cut in a slit-like shape by a set of slits as shown in figure~\ref{2Dbkg}. 

The detector was placed at its maximum distance possible from the sample position, $G$ in figure~\ref{amor}, in order to minimize the effect of introducing an absorber (a Cadmium slab) in the beam of the sample position. With this absorber the main beam was meant to be blocked far from the detector without altering the environmental background at the instrument with the shutter open. Figure~\ref{2Dbkg} shows the reconstructed image of the detector (integrated over the full neutron wavelength range 2.5-15\AA) when the absorber is out the beam (beam transmitted) and when it is in the beam (beam blocked). The absorber has the effect of decreasing the transmitted beam by approximately 3 orders of magnitude, while leaving the neutron background that reaches the detector unaltered. 

\begin{figure}[htbp]
	\centering
	\includegraphics[width=1\textwidth,keepaspectratio]{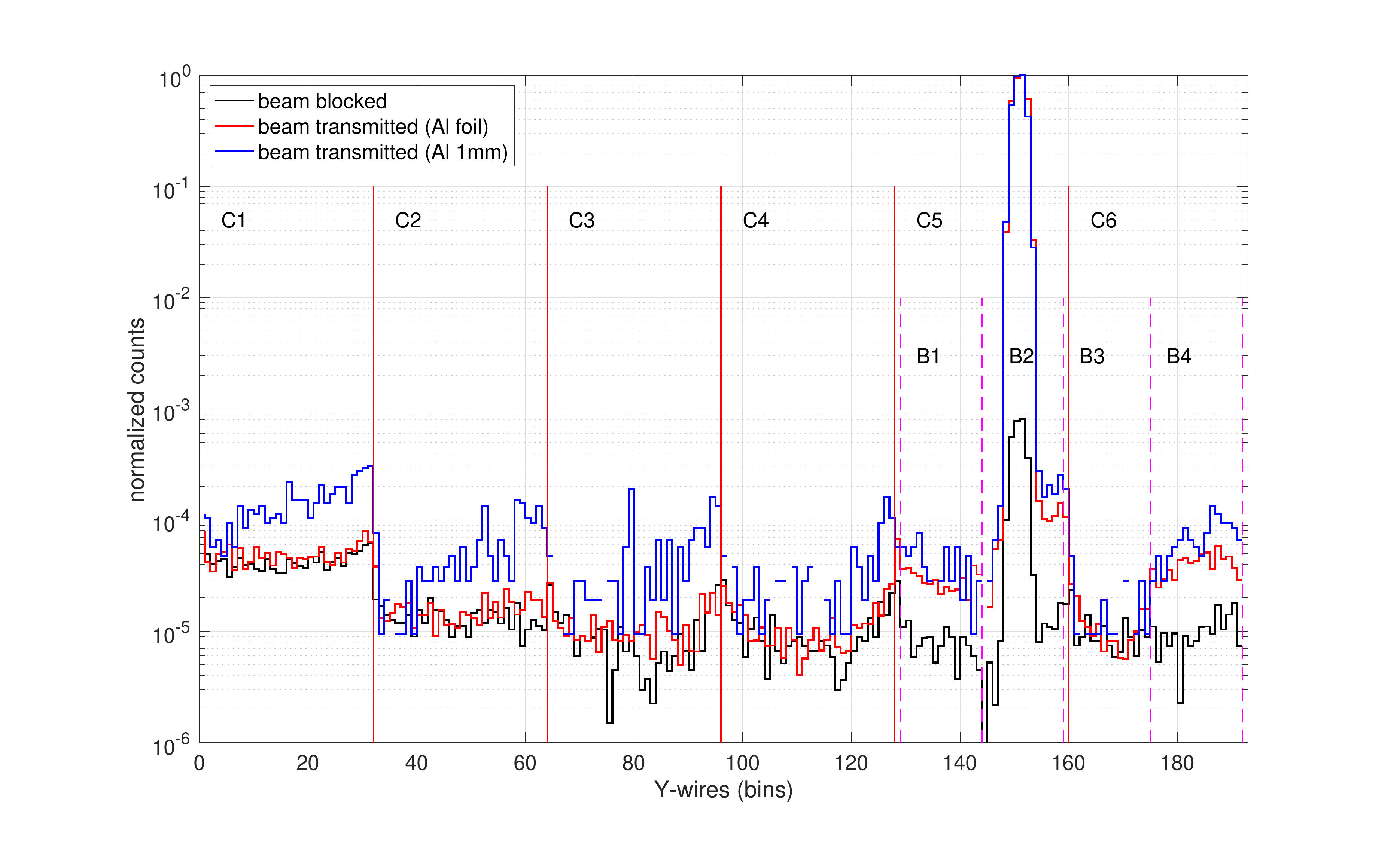}
	\caption{\label{1DBkg} \footnotesize Normalized counts in the 6 cassettes from the 2D image in figure~\ref{2Dbkg} integrated over the X-direction (strips) and integrated over the full neutron wavelength range 2.5-15\,\AA\,for three different configurations: beam blocked at the sample position (black), and beam transmitted with a 50 $\mu$m thick Al foil (red) and 1\,mm Al plate as entrance window of the detector. Each wire bin on the horizontal axis corresponds to approximately 0.35\,mm. Each curve is normalized to the maximum intensity in bin 151.}
\end{figure}

Figure~\ref{1DBkg} shows the normalized projection, against the wire axis, of the detector image (in figure~\ref{2Dbkg}) by summing over the strips in three configurations: when the beam is blocked (black curve), and when the beam is transmitted (red and blue curves). Each curve is normalized to the maximum intensity in bin 151. The six cassettes (C1 to C6) and four other regions (B1 to B4) are highlighted for ease of the following discussion. 

\begin{figure}[htbp]
	\centering
	\includegraphics[width=0.7\textwidth,keepaspectratio]{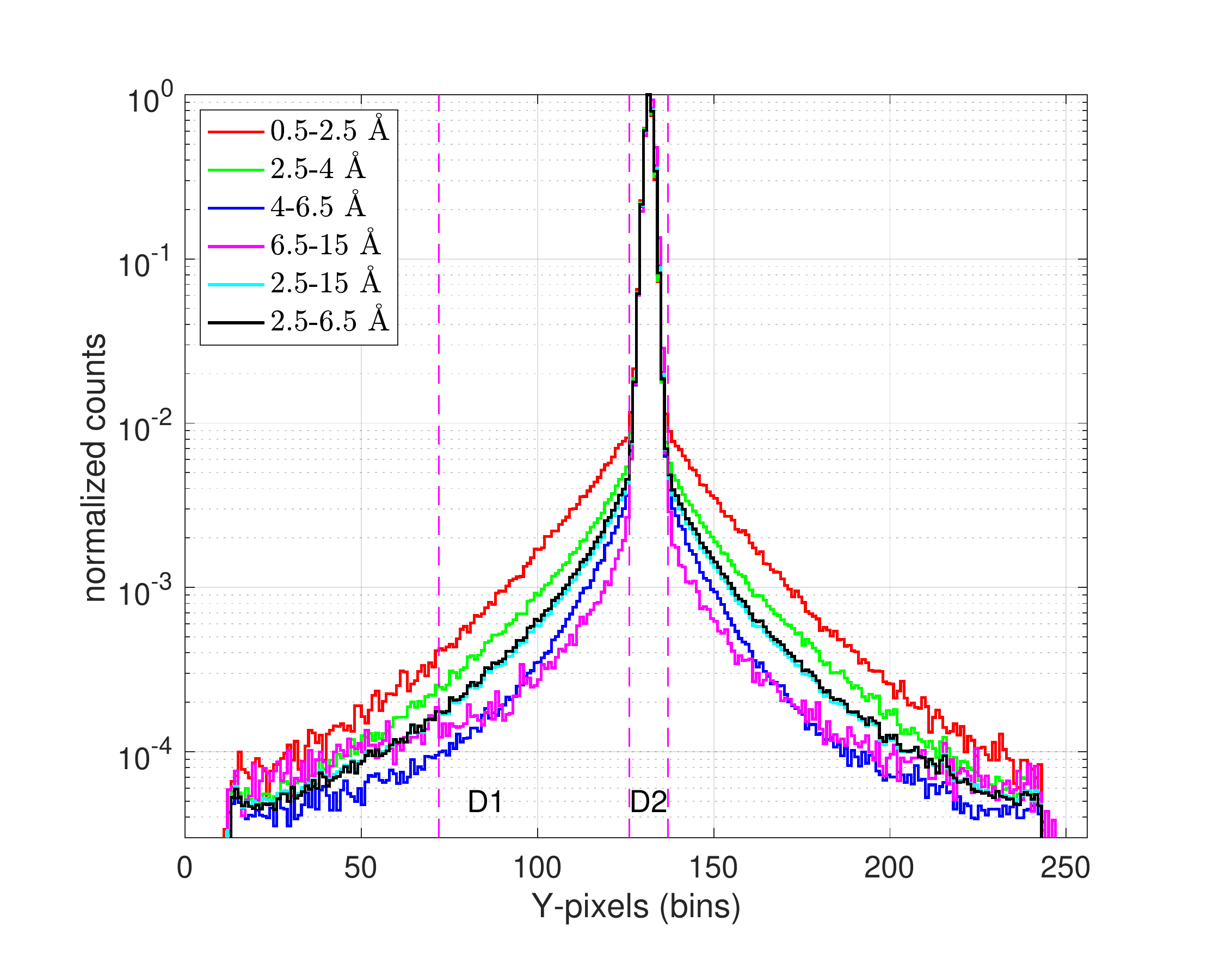}
	\caption{\label{BkgHe} \footnotesize Intensity distribution of the He-3-based AMOR detector integrated horizontally. Each bin on the horizontal axis corresponds to approximately 0.7\,mm. Each curve is normalized to the maximum intensity in bin 131.}
\end{figure}  

Since the Multi-Blade is operated at atmospheric pressure, a thin Al foil can be used as an entrance window. A comparison between a 1\,mm-thick Al window and a $\approx 50\, \mu$m Al foil is also shown in figure~\ref{1DBkg} (red and blue curves) when the beam is transmitted. The advantage of using a thin foil is immediately clear; the background generated by the scattering at the entrance window far from the direct beam appears lower (red curve against blue curve) and comparable to the environmental background with blocked beam (black curve) by almost an order of magnitude. Note that He-3-based neutron detectors for reflectometry are usually operated at a pressure of several bars and the entrance window (generally Al) is consequently a few millimeters thick. The background generated by the scattering on such a thick window is therefore much larger than that can be caused by a foil. In~\cite{INSTR_ESTIA} it has been shown that this background can reach approximately $10^{-3}$ peak to tail. A similar test was performed with a slit-like neutron beam impinging on the He-3-based neutron detector at AMOR. Figure~\ref{BkgHe} shows the normalized intensity of the AMOR detector integrated horizontally (over X) for various ranges of neutron wavelength. Each curve is normalized to the maximum intensity in bin 131.

Referring to figure~\ref{1DBkg}, when the beam is blocked the direct beam is attenuated by 3 orders of magnitude. By comparing the red and the black curves the background level far from the direct beam (C1 to C4) is similar, i.e. the contribution to the background due to the spurious scattering happening at the blade is similar to the environmental background. Referring to the results obtained at the CRISP reflectometer at ISIS with a 4.4 $\mu$m-thick $^{10}$B$_{4}$C coating~\cite{MIO_MB16CRISP_jinst}, the background generated by the scattering from the titanium substrate is reduced by approximately one order of magnitude, from $10^{-4}$ to $10^{-5}$ with the nominal coating thickness $>7.5\mu$m used in the present detector.

If the beam is transmitted, the background, close to the direct beam, B1, B4 in figure~\ref{1DBkg}, is higher than the environmental one (black curve). Note that this background is less affected by the thickness of the window. This effect is shown in the difference between the blue and red curves away from the direct beam peak (C1 to C4). The front of a cassette is more exposed to unwanted events generated by scattered neutrons, e.g. from the entrance window, because of the detector geometry. This occurs in a slight slope along the wire plane when the 1 mm-thick Al window is employed, blue curve in figure~\ref{1DBkg}.    

This background increase might be attributed, instead, to the extension of the beam and a possible reflection/scattering of the main beam on the edges of the collimation slits. The tail of the beam profile should be recorded in the wires in cassette C6 adjacent to the peak as observed in B1. On the contrary, the counts detected with the beam transmitted and beam blocked are comparable. In B4 the counts increase again, but the difference between the two Al windows employed is negligible. The similar intensities between region B1 and B4 may suggest that this is not a scattering from the substrate. It can be attributed to some reflection of the direct beam with the collimation slits.

Most of the homogeneous background originates from the fast neutrons in the SINQ instrument hall and cosmic rays, and therefore independent of the AMOR incident beam. All cassettes show background from fast neutrons but, from figures~\ref{1DBkg} and~\ref{2Dbkg}, the cassette at top (C1) shows a higher background with respect to the other cassettes. This does not change between the blocked (black) and transmitted beam (red) configurations. Fast and epithermal neutrons coming from the target at PSI are scattered back from the ceiling and thermalized by the detector vessel which is only shielded with a 5\,mm Al and a 2\,mm-thick Mirrobor layer. Both also moderates fast neutrons. This effect is not affecting the other cassettes that are instead shielded by the one on top. 

In order to compare the results obtained with the Multi-Blade at CRISP~\cite{MIO_MB16CRISP_jinst} and at AMOR and the He-3-based AMOR detector, a Figure-of-Merit (FoM) is defined, in equation~\ref{fomdefi}, as the integral of the counts in the peak where the beam is impinging over the mean of counts in an adjacent region of the detector. 

\begin{equation}\label{fomdefi}
FoM_{MB} = \frac{\int_{B2}P_{MB}\,dY}{<\overline{P_{MB}}>_{(C2,C3,C4,B1)}}, \qquad FoM_{He3} = \frac{\int_{D2}P_{He3}\,dY}{<\overline{P_{He3}}>_{(D1)}}
\end{equation}

Where $P_{x}$ is the projection of the detector image integrated over the horizontal direction (X) and shown in figures~\ref{1DBkg} and~\ref{BkgHe}.  For both CRISP and AMOR the counts in the peak are taken in the region B2 (see figure~\ref{1DBkg}) and for the He-3 detector in the region D2 (see figure~\ref{BkgHe}). The mean ($<\overline{P_{x}}>_{(k)}$) outside the direct beam intensity is taken in the three adjacent cassettes (C2 to C4) and part of cassette 5 (region B1) for the Multi-Blade and in region D1 for the He-3 detector. The same area is compared in the two detectors. Each wire bin on the horizontal axis corresponds to approximately 0.35\,mm for the Multi-Blade and 0.7\,mm for the He-3 detector. Regions C1 to C4 and B1 (figure~\ref{1DBkg}), or D1 (figure~\ref{BkgHe}) is approximately 38\,mm. The cassettes 1 (C1) and 6 (C6) are omitted in the calculation for what it has been stated above about a higher background in the cassette 1 due to faster neutrons and some spurious reflection of the direct beam with the collimation slits. 

The FoM calculated for several ranges of neutron wavelengths are shown in table~\ref{tab14} for a 1\,mm Al-window used on the Multi-Blade at CRISP and AMOR and Al-foil used at AMOR and for the He-3-based AMOR detector. Note that the wavelength ranges have been chosen accordingly the neutron wavelengths that will be used at the two ESS reflectometers, for which 4\,\AA\, and 2.5\,\AA\,are the shortest wavelengths available at Estia~\cite{INSTR_ESTIA,INSTR_ESTIA0,INSTR_ESTIA1,INSTR_ESTIA2} and FREIA~\cite{INSTR_FREIA,INSTR_FREIA2} respectively. 6.5\,\AA\, is the largest neutron wavelength available at CRISP.

\begin{table}[htbp]
	\centering
	\caption{\label{tab14} \footnotesize Figure-of-Merit (FoM) ($\times 10^4$) as the integral of the counts in the direct beam peak over the mean of counts in an adjacent region of the detector.}
	\smallskip
	\begin{tabular}{|l|c|c|c|c|c|c|}
		\hline
		\hline
		 FoM ($\times 10^4$)& \multicolumn{6}{|c|}{\bfseries wavelength range (\AA)}\\ 
		 &   0.5-2.5 & 2.5-4 & 4-6.5 & 2.5-6.5 & 6.5-15 & 2.5-15 \\ 
		\hline
		MB@CRISP~\cite{MIO_MB16CRISP_jinst} (4.4$\mu m$ coating) &0.4 & 4.0 & 5.2 & 5.2 & n/a & n/a \\
		 \hline
		MB@AMOR  (>7.5$\mu m$ coating) & & & & & &\\
		1mm Al window &n/a & 7.3 & 8.0  & 7.4 & 7.2 & 7.3 \\ 
		Al foil window  & n/a & 22.6 & 47.7 & 40.5 & 11.3 & 21.1 \\ 
		\hline
		He3@AMOR  & 0.1 & 0.2 & 0.5 & 0.3 & 0.8 & 0.4 \\ 
		\hline
		\hline
	\end{tabular}
\end{table}

The larger the FoM, the better is the signal-to-background ratio for a given configuration. It indicates how many orders of magnitude the signal exceeds the average background. From table~\ref{tab14}, any configuration for the Multi-Blade at either CRISP or AMOR is performing better than the He-3 based detector. The latter detector has a entrance window of several mm. The Multi-Blade with 4.4\,$\mu$m coating tested at CRISP had a 1\,mm Al-window as the present detector. By comparing the two the effect of having a thicker $\mathrm{^{10}B_4C}$ coating (>7.5\,$\mu$m) improves the detector performance by a factor up to 2. The latter is even improved further (of about a factor 3 over the whole wavelength range, 2.5-15~\AA) when the 1\,mm Al-window is replaced with the Al-foil window.

\subsection{Uniformity performance and efficiency correlation}\label{parax_unif}
A uniform response is of great importance to ensure the optimal performance of the Multi-Blade detector. In the case of a modular design, e.g. the Multi-Blade, a high mechanical precision is required in order to minimize differences between the units, which leads to intrinsic dis-uniformities, and in the data handling, when small adjustments can be applied to obtain a subsequent accurate data processing. 

\begin{figure}[htbp]
\centering
\begin{tabular}{cc}
\includegraphics[width=0.45\textwidth,keepaspectratio]{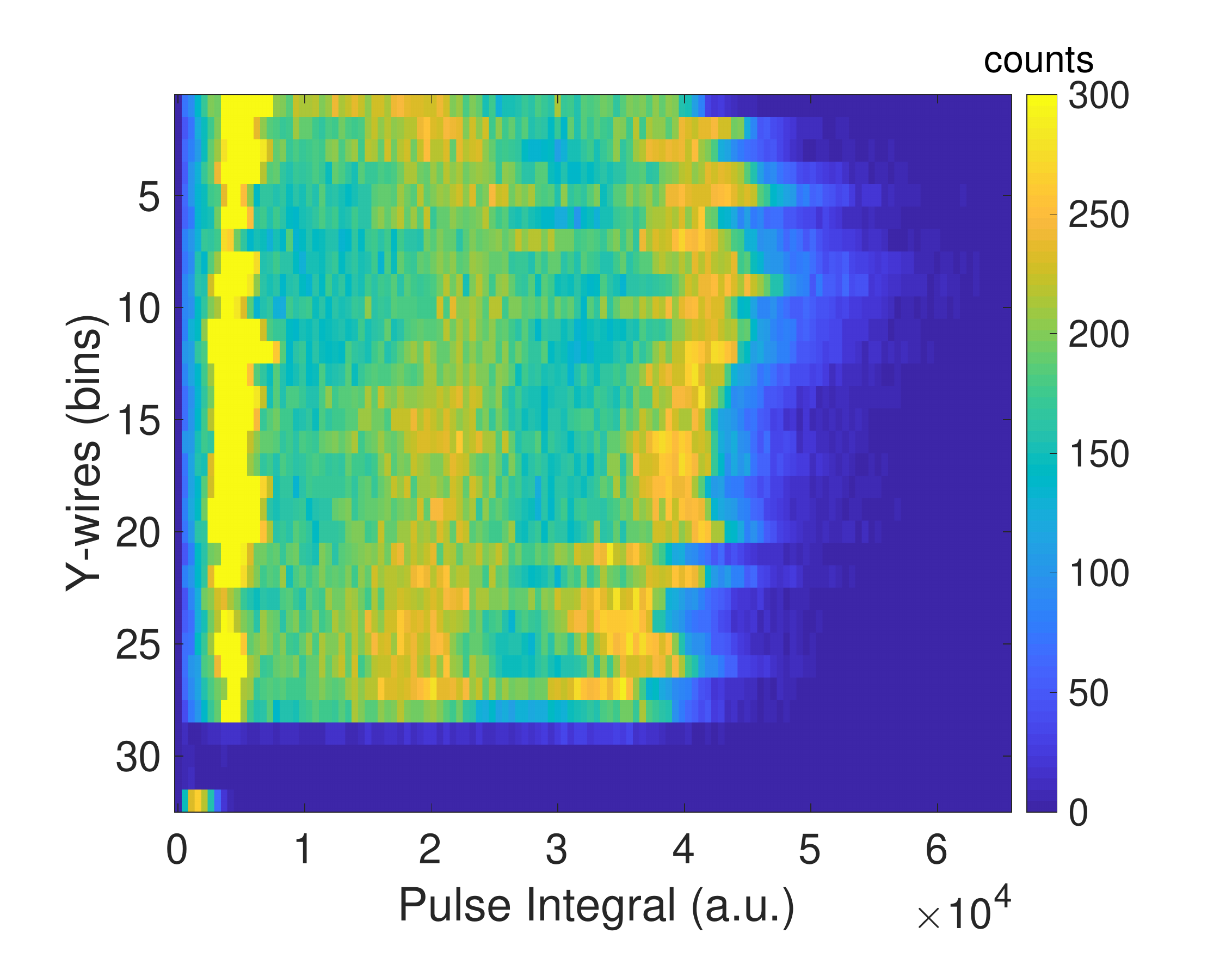}&
\includegraphics[width=0.45\textwidth,keepaspectratio]{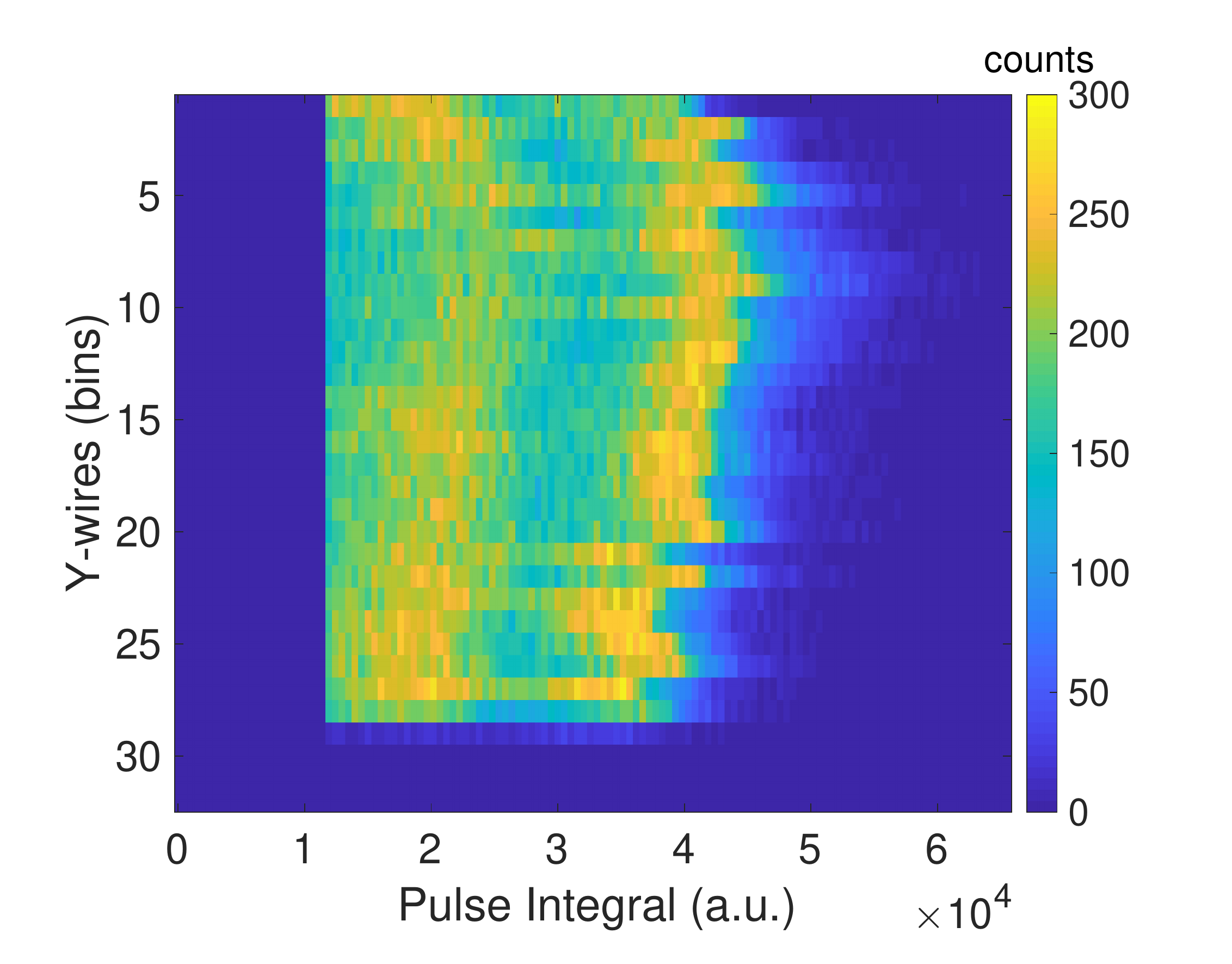}\\
 (a) & (b) \\
\includegraphics[width=0.45\textwidth,keepaspectratio]{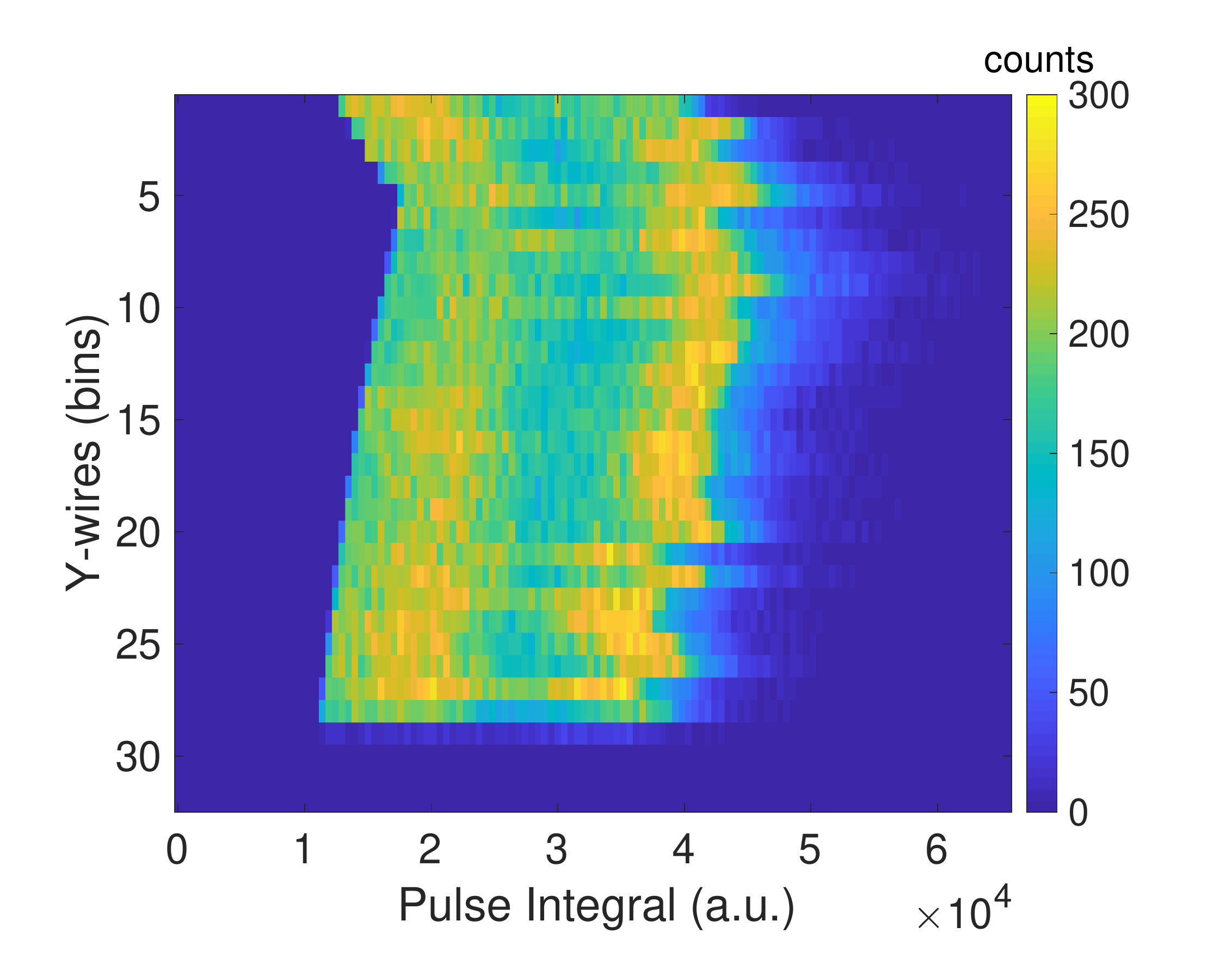}&
\includegraphics[width=0.45\textwidth,keepaspectratio]{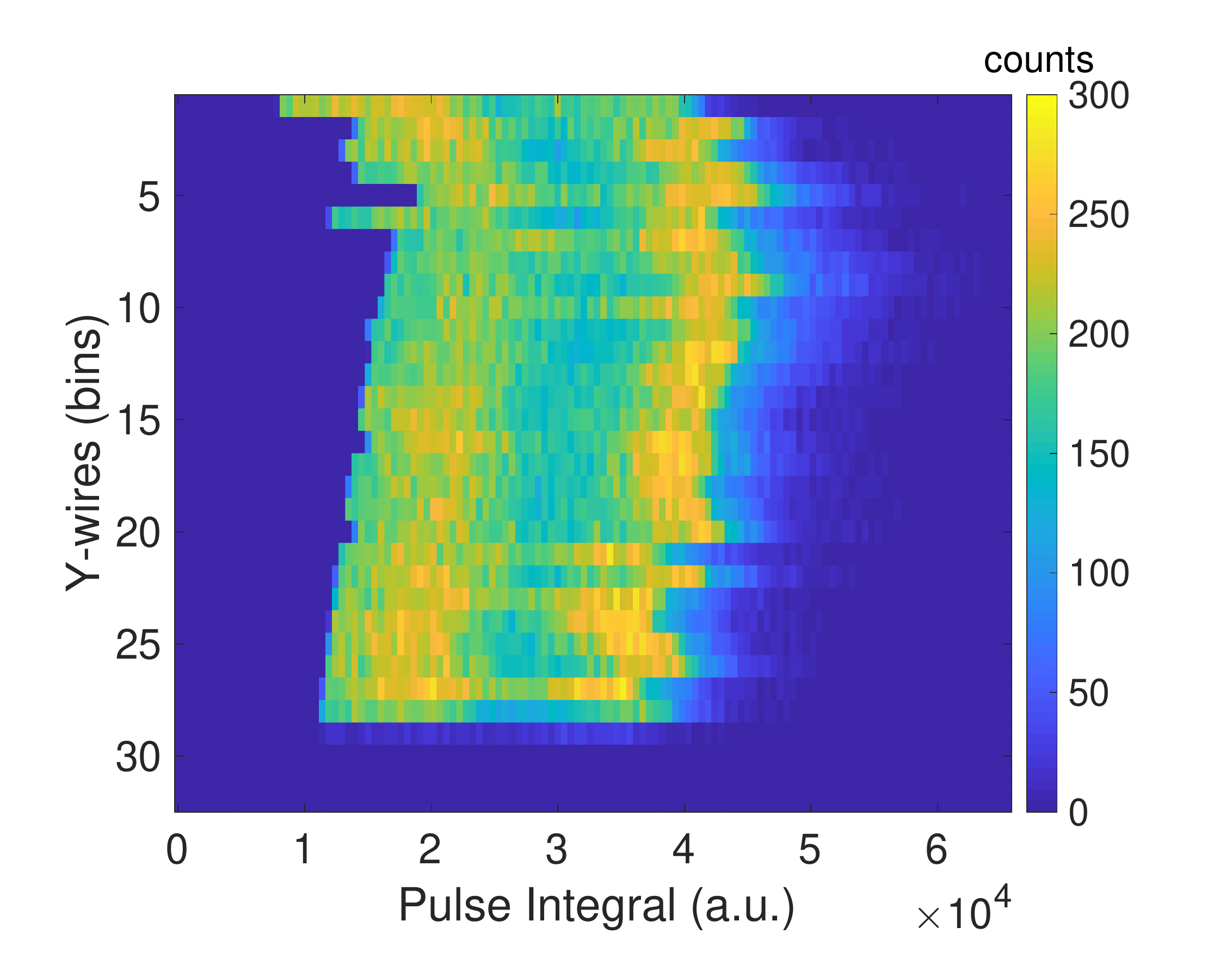}\\
(c) & (d) 
\end{tabular}
\caption{\label{Unifphs} \footnotesize PHS for each wire of one of the cassettes of the Multi-Blade. The neutron conversion fragment peaks ($Li$ and $\alpha$ particles, at $\approx 2\cdot10^4$ and $\approx 4\cdot10^4$ on the horizontal axis respectively) are visible. Wires from 29 to 32 are shadowed by the neighbouring cassettes. (a) No software threshold applied, only hardware thresholds. (b) A common threshold is applied to all wire channels. (c) A staircase threshold is applied to the wire channels of the cassette following the gain variation due to the detector geometry. (d) A channel-by-channel threshold is applied to the wire channels matching their individual gain.}
\end{figure}   

As described in section~\ref{mbsetup}, each Multi-Blade module is an independent MWPC with a 2-D readout system. The inclined geometry of the cassettes gives a characteristic response of the electric field along the wire plane~\cite{MIO_MB2017}. This is reflected in a different gas gain for each wire. The most evident effect is a decrease in the gain for the first wires. Due to the design of the blade of the cassette, the anode to cathode gap increases towards the edge of the MWPC, as illustrated in figure~\ref{mbSketch}. Due to the arrangement over a circle, the units are not parallel but fixed with a relative angle of $0.14^{\circ}$. A slight decrease of the gas gain is, therefore, expected across the wire plane, further details and the calculation of the electric field can be found in~\cite{MIO_MB2017}. 
Figure~\ref{Unifphs} shows the PHS for each wire of one cassette when (b) a common threshold is applied (c) when a staircase threshold following the shape of the electric field is used and (d) when gain matching thresholds wire by wire is applied. Due to the anode-cathode distance variation, the gain increased from wires 1 to 4 and reaches its maximum at wire number 5, then it decreases again till the end of the wire plane (wire 32). In each PHS on each wire the two bright intensities corresponds to the full energy deposition of the Li and $\alpha$ particles (from the neutron conversion reaction) respectively. 

As mentioned in section~\ref{mbsetup}, by applying a software threshold it is possible to discriminate against background events, i.e. gamma rays~\cite{MIO_MB16CRISP_jinst} visible in figure ~\ref{Unifphs} (a). These generate a smaller charge with respect to the neutron conversion fragments (Li and $\alpha$ particles), therefore smaller amplitudes are recorded, that can be easily neglected by applying a common threshold (in the software), figure~\ref{Unifphs} (b).  

\begin{figure}[htbp]
	\centering
	\includegraphics[width=1\textwidth,keepaspectratio]{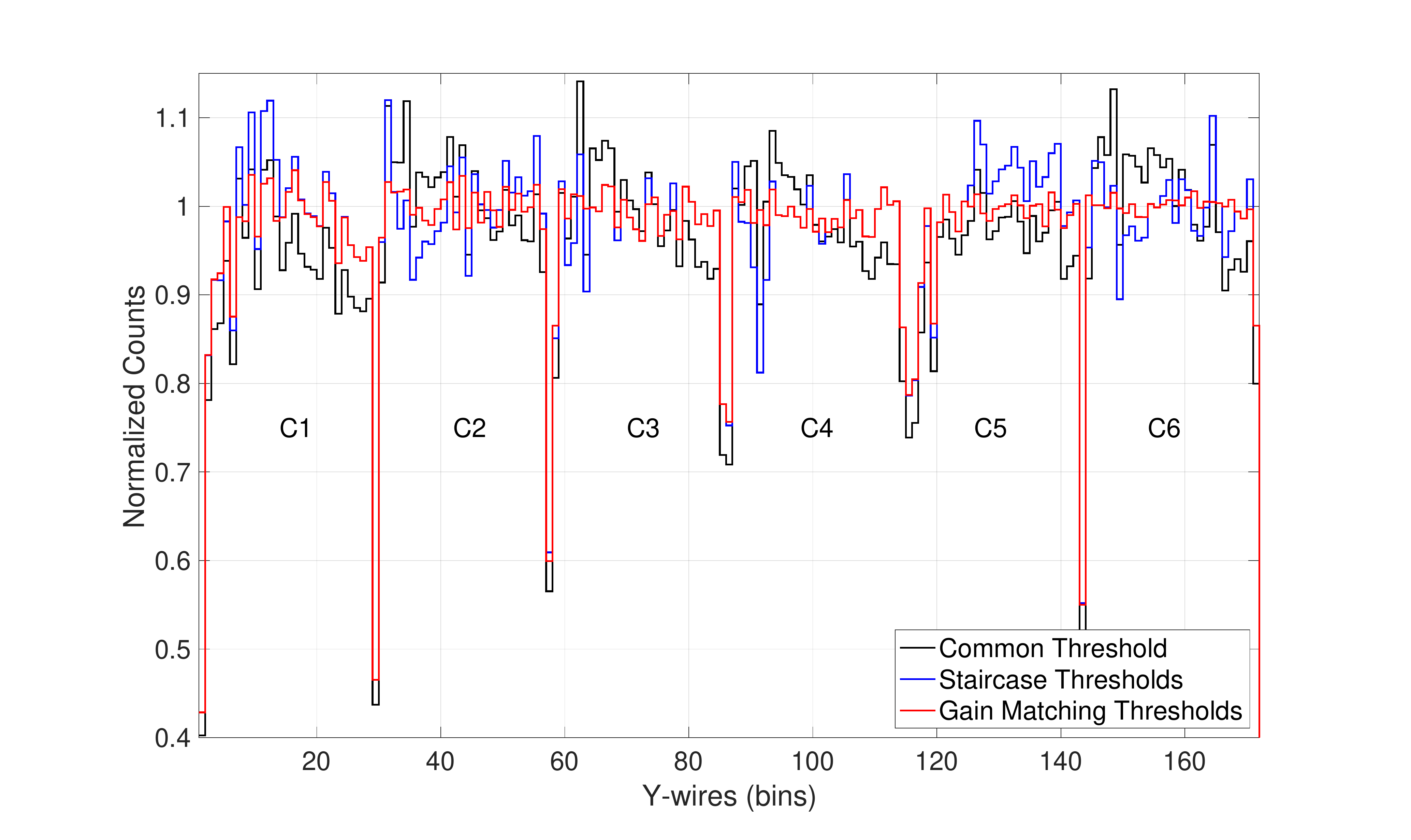}
	\caption{\label{Unif1d} \footnotesize Uniformity plot, 1D projection of the strips over the wires for the six cassettes of the Multi-Blade. Uniformity obtained with a common threshold is applied to all wire channels, a staircase threshold is applied to the wire channels of the cassette following the gain variation due to the detector geometry and a channel by channel threshold is applied to the wire channels matching their individual gain (respectively b, c and d in figure~\ref{Unifphs}).}
\end{figure}

\begin{figure}[htbp]
	\centering
	\includegraphics[width=1\textwidth,keepaspectratio]{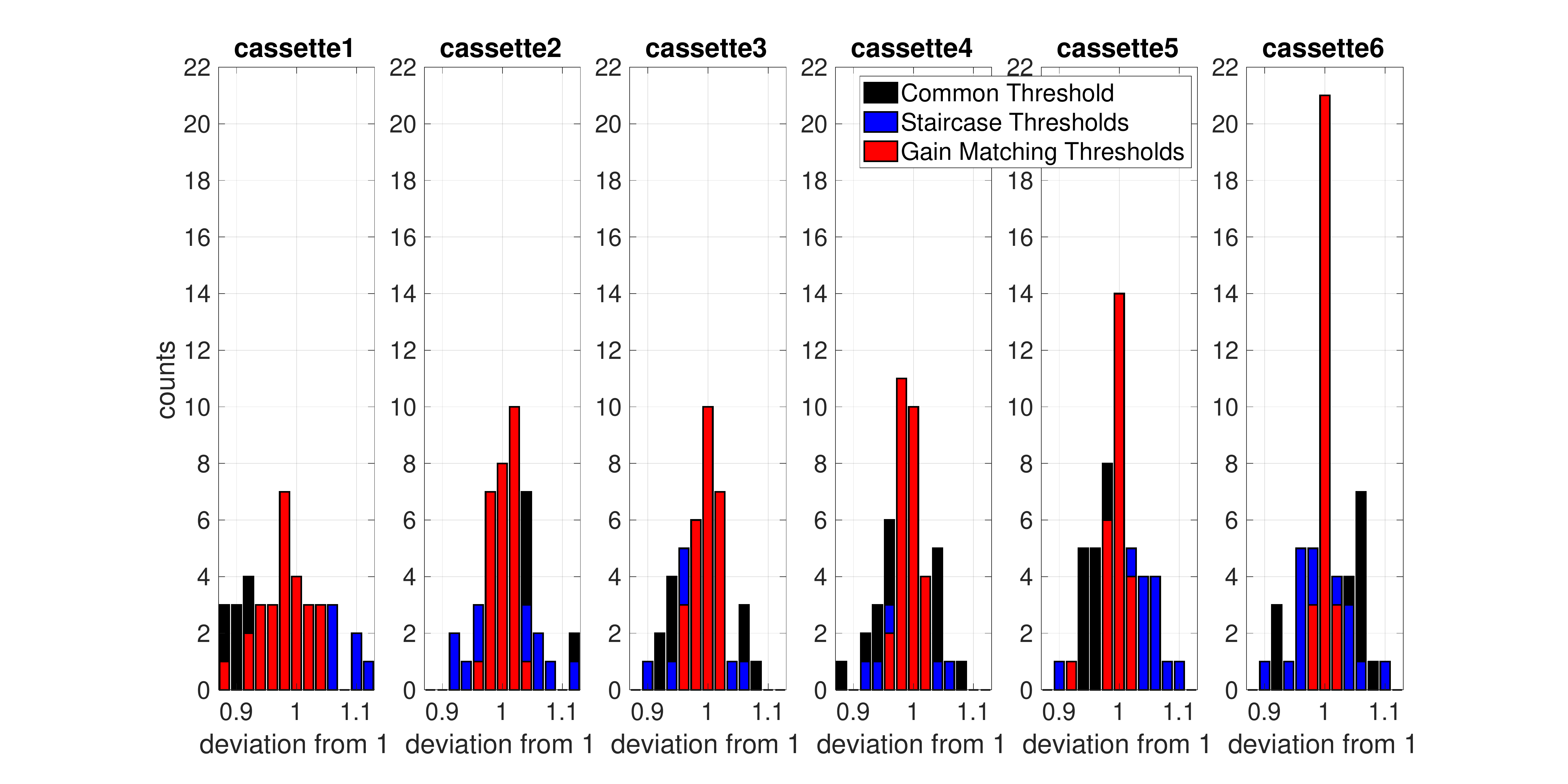}
	\caption{\label{Unif1ddistri} \footnotesize Distribution of the normalized counts in figure~\ref{Unif1d} given per cassette and for the three cases of applied thresholds (cases b, c and d). Each bin on the horizontal axis is 2\%.}
\end{figure}

In order to improve the uniformity of the detector, a further tuning can be done as shown in figure~\ref{Unifphs} (c) and (d). A staircase threshold can be applied to match the shape of the electric field, or individual thresholds can be chosen to match the gain of the different wires. The uniformity for the six cassettes of the detector is shown in figure~\ref{Unif1d}. This plot is the projection on the wire axis by summing over the strips integrated over the full ToF range. Note that the shadowed channels (where the cassette overlap) are removed in the plot and the visible gaps between the cassettes are a physical effect due to the charge collection at the first wire. This effect has been discuses in details in~\cite{MIO_MB16CRISP_jinst}, and the efficiency drop in these gaps never exceeds $\approx$50\%. Figure~\ref{Unif1ddistri} shows the distribution of the normalized counts on each wire bin in figure~\ref{Unif1d} for each cassette independently and for the three cases of applied thresholds (cases b, c and d).

A uniformity less than $\pm 5 \%$ can be achieved when the gain matching thresholds are applied, red curve in figure~\ref{Unif1d}. In the case of a staircase threshold, blue line, the uniformity is about $\pm 8 \%$. It can be noticed that the intrinsic non-uniformity of each module, due to the hardware components, varies the accuracy of the applied cut. Due to mechanical problems the cassettes one and five were expected to show a worse uniformity compared to the others. Cassettes two, three, four and six indeed are more uniform and do not exceed the $\pm 4 \%$ uniformity. The cassette six is the more uniform and shows a uniformity within $\pm 2\%$.
On the other hand, when a common threshold is applied, a uniformity around 10\% is obtained. Nevertheless, it is enough to neglect the background events. With respect to the previous characterization~\cite{MIO_MB16CRISP_jinst} a factor two or better uniformity is observed. 

When using the high intensity focusing mode, described in section~\ref{sec:estia}, the divergence of the beam allows to cover large areas of the detector for a single measurement. The uniformity is, therefore, an important feature in order to distinguish weak scattering signals, especially for off-specular reflectometry experiments. It is therefore crucial to implement a channel-by-channel calibration of the thresholds to get the desired uniformity within 2\% along with a better mechanical precision which will ensure a smaller unit to unit variation and across a single cassette.

A further effect which affects the Multi-Blade performance is the angular uniformity. The efficiency of the detector is a function of the incidence angle of neutrons on the $\mathrm{^{10}B_4C}$-layer surface (the angle $\beta$ in figure~\ref{mbSketch}). The nominal value for $\beta$ is 5$^{\circ}$, however both the extension of the sample and the varying distance between front and back of the blade, vary the incidence angle locally, leading to a variation of the efficiency. Note that the length of a blade (that front to back is approximately 130\,mm) has been optimized to minimize this effect, and as this is calculable, it can be corrected in the data.

\begin{figure}[htbp]
	\centering
	\includegraphics[width=0.7\textwidth,keepaspectratio]{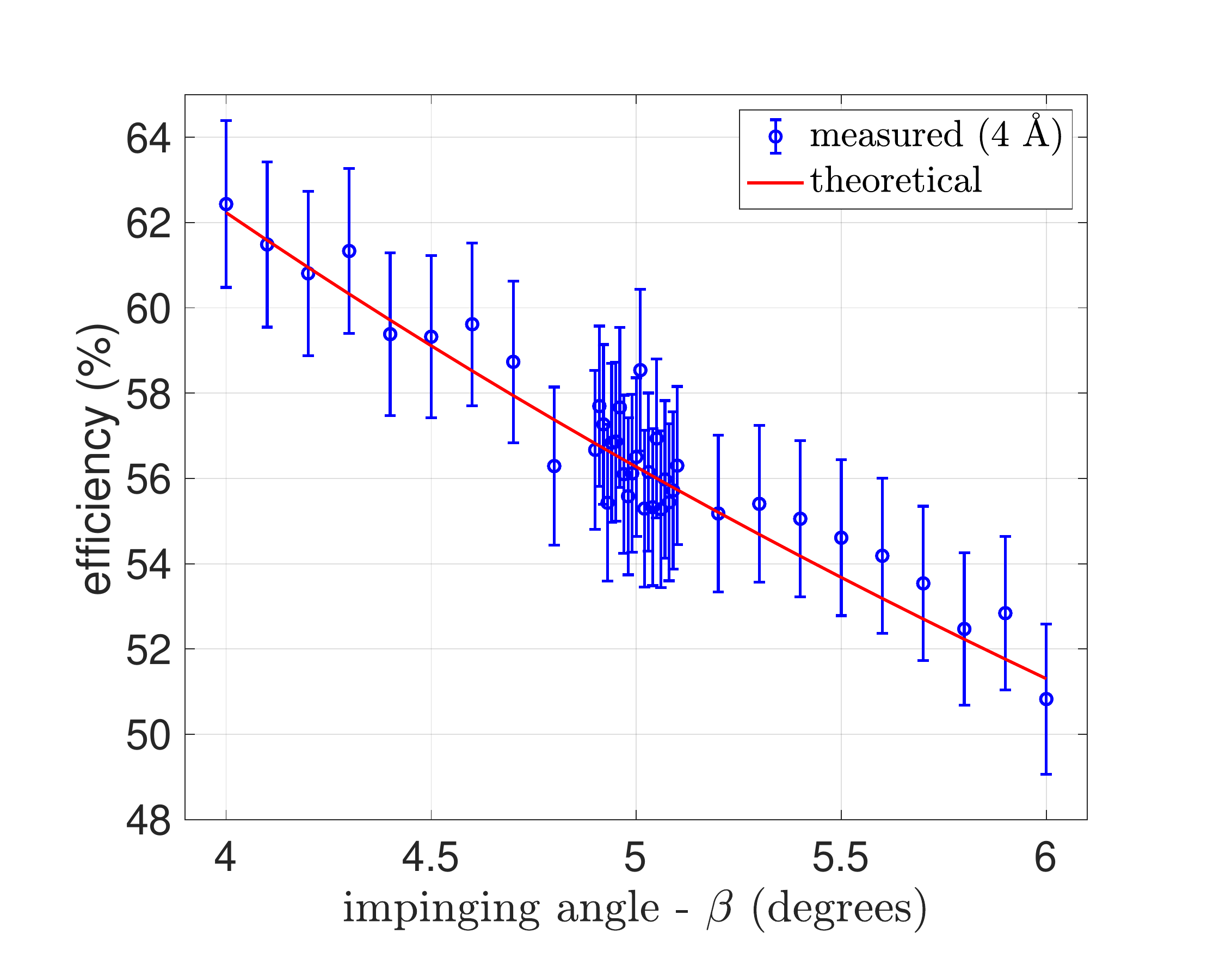}
	\caption{\label{amorScan} \footnotesize The efficiency variation (gating the neutron wavelengths around the neutron wavelength peak at AMOR, 4\AA) as a function of the impinging angle ($\beta$) of the neutrons on the blade coated with a $\mathrm{^{10}B_4C}$-layer (blue) and the calculated theoretical efficiency~\cite{MIO_analyt} (red).}
\end{figure}

A set of measurements was performed to investigate this effect. The incident beam has been centred in the middle of a cassette, this serves as centre of rotation of the detector with respect to the focal point, $F$ in sketch~\ref{amor}. An angular scan have been performed for $\beta = 5^{\circ} \pm 1^{\circ}$ around the pivoting point. Angles between $4.9^{\circ}$ and $5.1^{\circ}$ are scanned with a step of $0.1^{\circ}$. The efficiency variation (gating the neutron wavelengths around the neutron wavelength peak at AMOR, 4\AA) is shown in figure~\ref{amorScan} along with the calculated theoretical efficiency (red curve)~\cite{MIO_analyt}. 

The Multi-Blade is designed so that the projection of one cassette over $\beta$ is about $h = 11\,$mm (this value has been optimized to match the sample-detector distance foreseen at the Estia and the FREIA reflectometers~\cite{INSTR_ESTIA,INSTR_FREIA}.) Considering a point-like sample at a sample-detector distance of 4\,m (Estia), the angle subtended by one cassette toward the sample (pivoting center) is approximately $\approx 0.16^{\circ}$. An efficiency variation of $\pm 0.5\%$ is observed for an angular deviation within $\approx 0.16^{\circ}$, i.e. $\pm 0.08^{\circ}$. The region with a finer scan shown in figure~\ref{amorScan}. This is the maximum deviation allowed in a single blade.  
\\ If an extended sample is taken into account, the variation of $\beta$ at the center of the blade at highest reflection angle must be considered. Typical maximum sample extension along the beam direction for ESTIA will be $10\,mm$ and for FREIA $70\,mm$. At a reflection angle of 10$^{\circ}$, the variation of the incidence angle on the blades is $0.025^{\circ}$ and $0.17^{\circ}$ respectively. This is again within the $\pm 0.5\%$ efficiency variation. Note that a $\pm 0.5\%$ variation is well within the uncertainty of the measurement, i.e. such effect will not be discernible in the data.  

\subsection{Spatial resolution validation}
One of the key features that differentiates the Multi-Blade from the state-of-the-art technology detectors for neutron reflectometry applications is the enhanced spatial resolution. In the past years, detailed investigations have been carried out to prove and measure the spatial resolution of the detector~\cite{MIO_MB2014, MIO_MB2017}, based on the mutual information criterion, which was first applied and explained in~\cite{DET_patrickinfo}. A sub-millimeter spatial resolution, 0.54 mm across the wire plane, has been achieved with the Multi-Blade detector. The improvement of the $q-$resolution with the detector spatial resolution has been described in a previous work~\cite{MIO_ScientificMBcrisp}, together with the benefit in the data treatment and in the experimental procedure, e.g. time of measurements, driven by the Multi-Blade detector capabilities.

Although a dedicated measurement for spatial resolution was not performed, for completeness, a qualitative result is presented here, which is in agreement with the previous dedicated characterization. 

\begin{figure}[htbp]
	\centering
	\includegraphics[width=1\textwidth,keepaspectratio]{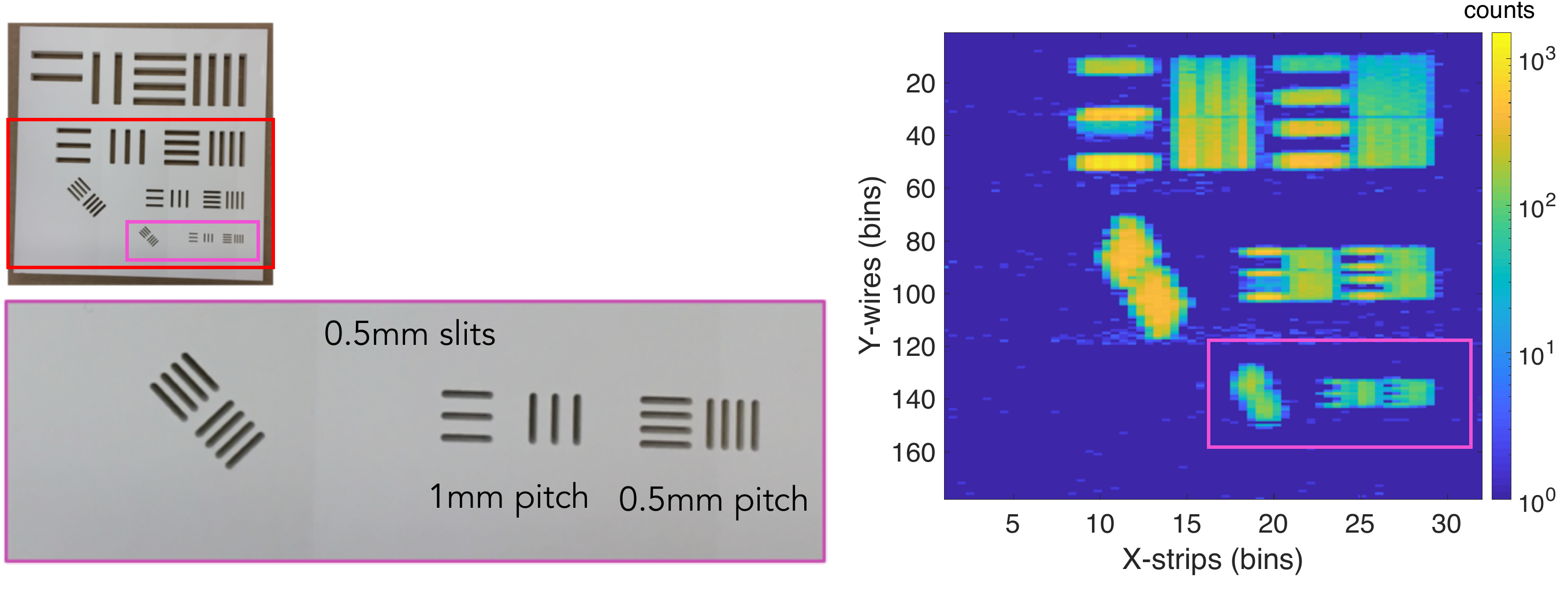}
	\caption{\label{maskmm} \footnotesize A picture of the Boron-Nitride (BN) mask (HeBoSint C100~\cite{hebosint}) (left) and the reconstructed image (right). The red box in the picture represents the full active area of the detector. In the zoomed picture are shown the 0.5 mm slits, both horizontal and vertical, with 1 mm and 0.5 mm pitches. The same area is highlighted by the purple box in the reconstructed image on the right. A bin in the vertical scale (wires) is 0.35\,mm and in the horizontal scale (strips) is 4\,mm.}
\end{figure}   

A set of measurements was performed using Boron-Nitride (BN) masks (HeBoSint C100 ~\cite{hebosint}) in front of the active area of the detector. The same masks have been employed to investigate the effect of reconstruction algorithms on position resolution, whose results are published in~\cite{MIO_MB16CRISP_jinst}. In figure~\ref{maskmm} both the picture of the mask and the raw reconstructed image from the measurements is shown. The red frame highlight the active area of the detector ($130 \times 60\,$mm$^2$), full image on the right, while the area enclosed in the purple box is shown is the zoomed picture and pointed out in the reconstructed image. Both the horizontal and vertical slits have a 0.5 mm wide opening, the pitch varies from 1 mm (left side) to 0.5 mm (right side). 

The horizontal slits are well distinguished across the wire plane (vertical direction, Y), both for the 1 mm and the 0.5 mm pitch, where a spatial resolution of $\approx 0.5\,$mm is expected. This result confirms that a sub-millimetre spatial resolution, about half a millimetre, can be achieved with the Multi-Blade detector. The resolution along the strip plane is about 3.5 mm, therefore the vertical slits cannot be distinguished. 

\section{Conclusions}
In the past years, continuous developments have been suggested to make progress in neutron reflectometry. One of the most appealing techniques is proposed by Estia~\cite{INSTR_ESTIA}, a forthcoming reflectometer at ESS~\cite{ESS}, which will exploit the focusing reflectometry technique to push forward the investigations held with such neutron scattering technique outside the core science case, enabling the possibility of better measurements. Along with it, better instrumentation is of a crucial importance for novel scientific fields.  

The Multi-Blade detector has been developed to meet the challenging requirements set by the new reflectometers at ESS. Numerous tests have been performed~\cite{MIO_MB2014,MIO_MB2017,
MIO_MB16CRISP_jinst, MIO_ScientificMBcrisp, MIO_fastn}, in these years to demonstrate the operation of this technology, to have a full characterisation, and to carry out reflectivity measurements.

The measurements performed at AMOR allow the exclusive possibility of combining an Estia-like instrument and the Multi-Blade detector, before ESS is in operation. Both the measurements exploiting the focusing mode and a characterization of the detector have been carried out, emphasising the impact the detector features have on the reflectivity measurements. 

The high-intensity specular reflectometry technique, operated thanks to the elliptic-shape guide (\textit{Selene}), was used to measure the reflectivity of a reference sample, a Ni/Ti multilayer. In the Estia proposal~\cite{INSTR_ESTIA}, the simulations of a sample with the same characteristics have been presented in order to demonstrate the concept of the method. A good agreement between the measurements and the simulations is shown in this manuscript. An indication of the future Estia experimental procedure and analysis is presented, i.e. by normalizing the reflected intensity to a reference sample with a high and well-known reflectivity, typically a supermirror.

The detector counting rate capability is the most demanding requirement set by the reflectometers amongst others. In most applications, indeed, a high intensity beam is focused to a small area on the detector. The flexibility of AMOR allows to perform specific measurements, varying both the incoming flux intensity and the illuminated area of the detector. A local instantaneous rate of approximately 3.4 kHz per detector pixel ($\approx 1\,$mm$^2$) is achieved, in agreement with the requirements set by Estia. Although this is the maximum measured rate without saturation of the detector, it is already more than one order of magnitude higher than the state-of-the-art detector technologies employed in neutron reflectometry.

A good signal-to-background separation is essential to measure a large dynamic range, thus detecting also the weakest signals from the scattering of a sample. A dynamic range of $10^{-7}$ is nowadays desired for future neutron reflectometers. An extensive investigation of the detector contribution to the background has been performed, by studying the scattering from the substrate and the effect of window thickness. With respect to the previous campaign of measurements~\cite{MIO_MB16CRISP_jinst}, the background of the detector has been improved by one order of magnitude, from $10^{-4}$ to $10^{-5}$. The latter is actually the intrinsic background present at the instrument without further detector shielding. Improvements can be made by enhancing the local shielding at the detector vessel.

Another important feature to consider when using the high-intensity reflectometry technique, it is the uniformity of the detector. A  divergent beam is used in this operation mode, and large areas of the detector may be illuminated, therefore a uniform response is required. The explanation of the detector principles and geometry is presented together with a software thresholds data processing implemented for the Multi-Blade. A uniformity better than 5\% can be achieved with the present detector, meeting the instrument requirements. Moreover, a factor two or better improvement is obtained with respect to the previous prototype. A study on the angular uniformity has been performed as well, in order to investigate the possible effects on the efficiency. A 0.5\% efficiency variation is observed for an angular divergence of $\pm 0.08^{\circ}$ on a point-like sample. An extended sample results in the same efficiency variation. This is below the uncertainty on the efficiency measurement. 

Further improvements have been accomplished with the present Multi-Blade detector, and the operation in an instrument like Estia has been successfully demonstrated. The final design for the installation at the ESS reflectometers is ready to be implemented.   

\acknowledgments This work was supported by the BrightnESS project (Horizon 2020, INFRADEV-3-2015, grant number 676548) and carried out as a part of the collaboration between the European Spallation Source (ESS - Sweden) and the Lund University (LU - Sweden).

The work originally started in the context of the collaboration between the Institut Laue-Langevin (ILL - France), the Link\"{o}ping University (LiU - Sweden) and the European Spallation Source (ESS - Sweden) within the context of the International Collaboration on the development of Neutron Detectors (www.icnd.org).

This work was partially performed at the Swiss Spallation Source SINQ at the Paul Scherrer Institute. The authors would like to thank the PSI detector group for the support during the tests and the PSI for the beam time on the AMOR instrument. 

Computing resources were provided by DMSC Computing Centre:\\ \small{\url{https://europeanspallationsource.se/data-management-software/computing-centre}}.

\bibliographystyle{ieeetr}
\bibliography{BIBLIOAmor}

\begin{thebibliography}{10}

\bibitem{MIO_MB2014}
F.~Piscitelli, J.~C. Buffet, J.~F. Clergeau, S.~Cuccaro, B.~Gu{\'e}rard,
  A.~Khaplanov, Q.~L. Manna, J.~M. Rigal, and P.~V. Esch, ``{Study of a high
  spatial resolution 10 B -based thermal neutron detector for application in
  neutron reflectometry: the Multi-Blade prototype},'' {\em Journal of
  Instrumentation}, vol.~9, no.~03, p.~P03007, 2014.

\bibitem{MIO_MB2017}
F.~Piscitelli, F.~Messi, M.~Anastasopoulos, T.~Bry{\'s}, F.~Chicken, E.~Dian,
  J.~Fuzi, C.~H{\"o}glund, G.~Kiss, J.~Orban, P.~Pazmandi, L.~Robinson,
  L.~Rosta, S.~Schmidt, D.~Varga, T.~Zsiros, and R.~Hall-Wilton, ``{The
  Multi-Blade Boron-10-based neutron detector for high intensity neutron
  reflectometry at ESS},'' {\em Journal of Instrumentation}, vol.~12, no.~03,
  p.~P03013, 2017.

\bibitem{MIO_MB16CRISP_jinst}
F.~Piscitelli, G.~Mauri, F.~Messi, M.~Anastasopoulos, T.~Arnold, A.~Glavic,
  C.~H{\"o}glund, T.~Ilves, I.~L. Higuera, P.~Pazmandi, D.~Raspino,
  L.~Robinson, S.~Schmidt, P.~Svensson, D.~Varga, and R.~Hall-Wilton,
  ``{Characterization of the Multi-Blade 10B-based detector at the CRISP
  reflectometer at ISIS for neutron reflectometry at ESS},'' {\em Journal of
  Instrumentation}, vol.~13, no.~05, p.~P05009, 2018.

\bibitem{MIO_ScientificMBcrisp}
G.~Mauri, F.~Messi, M.~Anastasopoulos, T.~Arnold, A.~Glavic, C.~H{\"o}glund,
  T.~Ilves, I.~Lopez~Higuera, P.~Pazmandi, D.~Raspino, L.~Robinson, S.~Schmidt,
  P.~Svensson, D.~Varga, R.~Hall-Wilton, and F.~Piscitelli, ``{Neutron
  reflectometry with the Multi-Blade 10B-based detector},'' {\em Proceedings of
  the Royal Society of London A: Mathematical, Physical and Engineering
  Sciences}, vol.~474, no.~2216, 2018.

\bibitem{MIO_HERE}
F.~Piscitelli, ``Novel boron-10-based detectors for neutron scattering
  science,'' {\em The European Physical Journal Plus}, vol.~130, no.~2,
  pp.~1--9, 2015.

\bibitem{MIO_MyThesis}
F.~Piscitelli, {\em Boron-10 layers, Neutron Reflectometry and Thermal Neutron
  Gaseous Detectors}.
\newblock PhD thesis, Institut Laue-Langevin and University of Perugia, 2014 -
  arXiv:1406.3133.

\bibitem{ILL}
``{ILL}.'' https://www.ill.eu.

\bibitem{ILL2}
G.~Kostorz, ``{Neutron scattering and materials research at the Institut Max
  von Laue-Paul Langevin},'' vol.~28, no.~2, pp.~61--73, 1976.

\bibitem{ILL3}
A.~Harrison, J.~L. Mart'nez, and R.~Wagner, ``{Renaissance: A Decade of
  Development at ILL},'' {\em Neutron News}, vol.~21, no.~2, pp.~11--14, 2010.

\bibitem{ESS}
``{European Spallation Source ESS ERIC}.'' http://europeanspallationsource.se.

\bibitem{ESS_TDR}
S.~Peggs. ESS Technical Design Report - (ESS-2013-0001).

\bibitem{ESS-design}
R.~Garoby, ``{The European Spallation Source Design},'' {\em Physica Scripta},
  vol.~93, no.~1, p.~014001, 2018.

\bibitem{ESS2011}
M.~Lindroos, S.~Bousson, R.~Calaga, H.~Danared, G.~Devanz, R.~Duperrier,
  J.~Eguia, M.~Eshraqi, S.~Gammino, H.~Hahn, A.~Jansson, C.~Oyon,
  S.~Pape-Moller, S.~Peggs, A.~Ponton, K.~Rathsman, R.~Ruber, T.~Satogata, and
  G.~Trahern, ``{The European Spallation Source},'' {\em Nuclear Instruments
  and Methods in Physics Research, Section B: Beam Interactions with Materials
  and Atoms}, vol.~269, no.~24, pp.~3258--3260, 2011.

\bibitem{INSTR_ESTIA}
J.~Stahn, ``Estia: A truly focusing reflectometer.'' ESS instrument proposal,
  2014.

\bibitem{INSTR_ESTIA0}
J.~Stahn, T.~Panzner, U.~Filges, C.~Marcelot, and P.~B{\"o}ni, ``Study on a
  focusing, low-background neutron delivery system,'' {\em Nuclear Instruments
  and Methods in Physics Research Section A: Accelerators, Spectrometers,
  Detectors and Associated Equipment}, vol.~634, no.~1, Supplement, pp.~S12 --
  S16, 2011.
\newblock Proceedings of the International Workshop on Neutron Optics NOP2010.

\bibitem{INSTR_ESTIA1}
{Stahn, J.}, {Filges, U.}, and {Panzner, T.}, ``Focusing specular neutron
  reflectometry for small samples,'' {\em Eur. Phys. J. Appl. Phys.}, vol.~58,
  no.~1, p.~11001, 2012.

\bibitem{INSTR_ESTIA2}
J.~Stahn and A.~Glavic, ``{Focusing neutron reflectometry: Implementation and
  experience on the TOF-reflectometer Amor},'' {\em Nuclear Instruments and
  Methods in Physics Research Section A: Accelerators, Spectrometers, Detectors
  and Associated Equipment}, vol.~821, no.~Supplement C, pp.~44 -- 54, 2016.

\bibitem{INSTR_FREIA}
H.~Wacklin, ``{FREIA: Reflectometer concept for fast kinetics at ESS - ESS
  instrument proposal}.''
  {https://europeanspallationsource.se/sites/default/files/freia\_proposal.pdf},
  2014.

\bibitem{INSTR_FREIA2}
H.~Wacklin, ``{Revealing Change Over Time, FREIA Brings Fast Kinetic Studies to
  Reflectometry}.''
  http://neutronsources.org/news/scientific-highlights/revealing-change-over-time-freia-brings-fast-kinetic-studies-to-reflectometry.html,
  2016.

\bibitem{DET_rates}
I.~Stefanescu, M.~Christensen, J.~Fenske, R.~Hall-Wilton, P.~Henry,
  O.~Kirstein, M.~M{\"u}ller, G.~Nowak, D.~Pooley, D.~Raspino, N.~Rhodes,
  J.~{\v S}aroun, J.~Schefer, E.~Schooneveld, J.~Sykora, and W.~Schweika,
  ``Neutron detectors for the {ESS} diffractometers,'' {\em Journal of
  Instrumentation}, vol.~12, no.~01, p.~P01019, 2017.

\bibitem{HE3S_kirstein}
O.~Kirstein and et~al., ``Neutron position sensitive detectors for the {ESS},''
  {\em arXiv:1411.6194}, vol.~Proceedings of the 23rd International Workshop on
  Vertex Detectors, 15-19 September 2014, Macha Lake, The Czech Republic,
  no.~PoS(Vertex2014)029, 2014.

\bibitem{INSTR_D17}
R.~Cubitt and G.~Fragneto, ``{D17: the new reflectometer at the ILL},'' {\em
  Applied Physics A}, vol.~74, no.~1, pp.~s329--s331, 2002.

\bibitem{INSTR_FIGARO}
R.~A. Campbell, H.~P. Wacklin, I.~Sutton, R.~Cubitt, and G.~Fragneto,
  ``{FIGARO: The new horizontal neutron reflectometer at the ILL},'' {\em The
  European Physical Journal Plus}, vol.~126, no.~11, pp.~1--22, 2011.

\bibitem{INSTR_R_Spin2}
J.~Major, H.~Dosch, G.~Felcher, K.~Habicht, T.~Keller, S.~te~Velthuis,
  A.~Vorobiev, and M.~Wahl, ``Combining of neutron spin echo and reflectivity:
  a new technique for probing surface and interface order,'' {\em Physica B:
  Condensed Matter}, vol.~336, no.~1, pp.~8 -- 15, 2003.
\newblock Proceedings of the Seventh International Conference on Surface X-ray
  and Neutron Scattering.

\bibitem{OTT_tiltof}
F.~Ott and A.~Menelle, ``{TilToF: A high-intensity space--time
  reflectometer},'' {\em Physica B: Condensed Matter}, vol.~385-386, pp.~985 --
  988, 2006.

\bibitem{OTT_gradtof}
F.~Ott and A.~de~Vismes, ``{RefloGrad/GradTOF: Neutron energy analysis for a
  very high-flux neutron reflectometer},'' {\em Physica B: Condensed Matter},
  vol.~397, no.~1, pp.~153 -- 155, 2007.

\bibitem{OTT_refocus}
F.~Ott and A.~Menelle, ``{REFocus: A new concept for a very high flux neutron
  reflectometer},'' {\em Nuclear Instruments and Methods in Physics Research
  Section A: Accelerators, Spectrometers, Detectors and Associated Equipment},
  vol.~586, no.~1, pp.~23 -- 30, 2008.
\newblock Proceedings of the European Workshop on Neutron Optics.

\bibitem{R_Cubitt1}
R.~Cubitt and J.~Stahn, ``Neutron reflectometry by refractive encoding,'' {\em
  The European Physical Journal Plus}, vol.~126, no.~11, pp.~1--5, 2011.

\bibitem{R_Cubitt2}
R.~Cubitt, H.~Shimizu, K.~Ikeda, and N.~Torikai, ``Refraction as a means of
  encoding wavelength for neutron reflectometry,'' {\em Nuclear Instruments and
  Methods in Physics Research Section A: Accelerators, Spectrometers, Detectors
  and Associated Equipment}, vol.~558, no.~2, pp.~547 -- 550, 2006.

\bibitem{AMOR_Clemens}
D.~Clemens, P.~Gross, P.~Keller, N.~Schlumpf, and M.~K{\"o}nnecke, ``{AMOR --
  the versatile reflectometer at SINQ},'' {\em Physica B: Condensed Matter},
  vol.~276-278, pp.~140 -- 141, 2000.

\bibitem{AMOR_Gupta}
M.~Gupta, T.~Gutberlet, J.~Stahn, P.~Keller, and D.~Clemens, ``{AMOR --- the
  time-of-flight neutron reflectometer at SINQ/PSI},'' {\em Pramana}, vol.~63,
  pp.~57--63, Jul 2004.

\bibitem{AMOR}
``{AMOR Instrument}.'' https://www.psi.ch/sinq/amor/amor.

\bibitem{PSI}
``{PSI}.'' {Paul Scherrer Institute (PSI) - https://www.psi.ch }.

\bibitem{PSI_SINQ}
W.~Wagner, Y.~Dai, H.~Glasbrenner, M.~Grosse, and E.~Lehmann, ``Status of sinq,
  the only mw spallation neutron source---highlighting target development and
  industrial applications,'' {\em Nuclear Instruments and Methods in Physics
  Research Section A: Accelerators, Spectrometers, Detectors and Associated
  Equipment}, vol.~562, no.~2, pp.~541 -- 547, 2006.
\newblock Proceedings of the 7th International Conference on Accelerator
  Applications.

\bibitem{FAC_BNC}
``Budapest neutron centre.'' http://www.bnc.hu.

\bibitem{FAC_BNC1}
L.~Rosta, ``Cold neutron research facility at the budapest neutron centre,''
  {\em Applied Physics A}, vol.~74, pp.~s52--s54, Dec 2002.

\bibitem{FAC_BNC2}
L.~Rosta and R.~Baranyai, ``Budapest research reactor -- 20 years of
  international user operation,'' {\em Neutron News}, vol.~22, no.~3,
  pp.~31--36, 2011.

\bibitem{CRISP1}
CRISP instrument manual 2010 -
  https://www.isis.stfc.ac.uk/Pages/crisp-instrument-manual-nov-2010.pdf, 2010.

\bibitem{ISIS}
ISIS Neutron and Muon Source - https://www.isis.stfc.ac.uk.

\bibitem{ISIS1}
C.~C. Wilson, ``Isis, the uk spallation neutron source - a guided tour,'' {\em
  Neutron News}, vol.~1, no.~1, pp.~14--19, 1990.

\bibitem{ISIS2}
C.~C. Wilson, ``A guided tour of isis - the uk spallation neutron source,''
  {\em Neutron News}, vol.~6, no.~2, pp.~27--34, 1995.

\bibitem{MIO_fastn}
G.~Mauri, F.~Messi, K.~Kanaki, R.~Hall-Wilton, E.~Karnickis, A.~Khaplanov, and
  F.~Piscitelli, ``{Fast neutron sensitivity of neutron detectors based on
  Boron-10 converter layers},'' {\em Journal of Instrumentation}, vol.~13,
  no.~03, p.~P03004 (arxiv:1712.05614), 2018.

\bibitem{SF1}
J.~Scherzinger, J.~Annand, G.~Davatz, K.~Fissum, U.~Gendotti, R.~Hall-Wilton,
  E.~H{\aa}kansson, R.~Jebali, K.~Kanaki, M.~Lundin, B.~Nilsson, A.~Rosborge,
  and H.~Svensson, ``{Tagging fast neutrons from an 241Am/9Be source},'' {\em
  Applied Radiation and Isotopes}, vol.~98, pp.~74 -- 79, 2015.

\bibitem{SF2}
F.~Messi, H.~Perrey, K.~Fissum, M.~Akkawi, R.~A. Jebali, J.~Annand, P.~Bentley,
  L.~Boyd, C.~Cooper-Jensen, D.~DiJulio, J.~Freita-Ramos, R.~Hall-Wilton,
  A.~Huusko, T.~Ilves, F.~Issa, A.~Jalgen, K.~Kanaki, E.~Karnickis,
  A.~Khaplanov, S.~Koufigar, V.~Maulerova, G.~Mauri, N.~Mauritzson, W.~Pei,
  F.~Piscitelli, E.~Rofors, J.~Scherzinger, H.~Soderhielm, D.~Soderstrom, and
  I.~Stefanescu, ``{The neutron tagging facility at Lund University},'' {\em
  arXiv:1711.10286 (submitted to IAEA Technical Report on Modern Neutron
  Detection (2017))}, 2017.

\bibitem{B4C_carina}
C.~H{\"o}glund, J.~Birch, K.~Andersen, T.~Bigault, J.-C. Buffet, J.~Correa,
  P.~van Esch, B.~Guerard, R.~Hall-Wilton, J.~Jensen, A.~Khaplanov,
  F.~Piscitelli, C.~Vettier, W.~Vollenberg, and L.~Hultman, ``{B4C thin films
  for neutron detection},'' {\em Journal of Applied Physics}, vol.~111, no.~10,
  2012.

\bibitem{B4C_carina3}
C.~H{\"o}glund, K.~Zeitelhack, P.~Kudejova, J.~Jensen, G.~Greczynski, J.~Lu,
  L.~Hultman, J.~Birch, and R.~Hall-Wilton, ``Stability of {10B4C} thin films
  under neutron radiation,'' {\em Radiation Physics and Chemistry}, vol.~113,
  pp.~14 -- 19, 2015.

\bibitem{B4C_Schmidt}
S.~Schmidt, C.~H{\"o}glund, J.~Jensen, L.~Hultman, J.~Birch, and
  R.~Hall-Wilton, ``Low-temperature growth of boron carbide coatings by direct
  current magnetron sputtering and high-power impulse magnetron sputtering,''
  {\em Journal of Materials Science}, vol.~51, no.~23, pp.~10418--10428, 2016.

\bibitem{MIO_analyt}
F.~Piscitelli and P.~V. Esch, ``Analytical modeling of thin film neutron
  converters and its application to thermal neutron gas detectors,'' {\em
  Journal of Instrumentation}, vol.~8, no.~04, p.~P04020, 2013.

\bibitem{EL_CAEN}
``{CAEN - Electronic Instrumentation}.'' http://www.caen.it.

\bibitem{MATLAB:R2019a_u3}
The Mathworks, Inc., Natick, Massachusetts, {\em {MATLAB version 9.6.0.1135713
  (R2019a) Update 3}}, 2019.

\bibitem{doublediscchopper}
A.~van Well, ``Double-disk chopper for neutron time-of-flight experiments,''
  {\em Physica B: Condensed Matter}, vol.~180-181, pp.~959 -- 961, 1992.

\bibitem{OTT_general}
F.~Ott and A.~Menelle, ``New designs for high intensity specular neutron
  reflectometers,'' {\em The European Physical Journal Special Topics},
  vol.~167, pp.~93--99, Feb 2009.

\bibitem{DMSC_Brigh}
``Brightneess deliverables - work package 5: Real-time management of {ESS}
  data.'' https://brightness.esss.se/archive/about/deliverables/, (2018).

\bibitem{essghal}
T.~Ghal, ``Hardware aspects, modularity and integration of an event mode data
  acquisition and instrument control for the european spallation source
  (ess),'' {\em Proceedings of ICANS XXI, Mito, Japan (2014), JAEA-Conf
  2015-002, http://dx.doi.org/10.11484/jaea-conf-2015-002 , arXiv:1507.01838},
  2014.

\bibitem{DMSC_EFU}
``Event formation unit source code (2018).''
  https://github.com/ess-dmsc/event-formation-unit, 2018.

\bibitem{DMSC_kafka}
``Apache kafka (2018).'' https://kafka.apache.org/.

\bibitem{DMSC_HDF}
``Hdf5 file format.'' https://www.hdfgroup.org.

\bibitem{DMSC_daquiri}
``Daquiri (2018).'' https://github.com/ess-dmsc/daquiri, 2018.

\bibitem{MB-DMSC}
M.~Christensen, M.~Shetty, J.~Nilsson, A.~Mukai, R.~A. Jebali, A.~Khaplanov,
  M.~Lupberger, F.~Messi, D.~Pfeiffer, F.~Piscitelli, T.~Blum, C.~S{\o}gaard,
  S.~Skelboe, R.~Hall-Wilton, and T.~Richter, ``Software-based data acquisition
  and processing for neutron detectors at european spallation
  source{\textemdash}early experience from four detector designs,'' {\em
  Journal of Instrumentation}, vol.~13, pp.~T11002--T11002, nov 2018.

\bibitem{DET_knoll}
G.~Knoll, {\em Radiation Detection and Measurement}.
\newblock John Wiley and Sons, Inc., third edition~ed., 2000.

\bibitem{g1}
{S. Agostinelli et al.}, ``{Geant4 ---- a simulation toolkit},'' {\em Nuclear
  Instruments and Methods in Physics Research A}, vol.~506, pp.~250--303, 2003.

\bibitem{MIO_MBscattsimu}
G.~Galg{\'{o}}czi, K.~Kanaki, F.~Piscitelli, T.~Kittelmann, D.~Varga, and
  R.~Hall-Wilton, ``Investigation of neutron scattering in the multi-blade
  detector with geant4 simulations,'' {\em Journal of Instrumentation},
  vol.~13, pp.~P12031--P12031, dec 2018.

\bibitem{DET_patrickinfo}
J.~F. Clergeau, M.~Ferraton, B.~Guerard, A.~Khaplanov, F.~Piscitelli, M.~Platz,
  J.~M. Rigal, T.~Daulle, and P.~V. Esch, ``An information-theoretical approach
  to image resolution applied to neutron imaging detectors based upon
  discriminator signals,'' {\em Proceedings of ANNIMA conference, Marseille
  2013, arXiv:1307.7507}, 2013.

\bibitem{hebosint}
``{HeBoSint Boron Nitride Components}.''
  http://www.henze-bnp.com/HeBoSint-boron-nitride-sintered-components.php.

\end{thebibliography}
\end{document}